\documentclass{article}

\usepackage{PRIMEarxiv}

\usepackage[utf8]{inputenc} 
\usepackage[T1]{fontenc}    
\usepackage{hyperref}       
\usepackage{url}            
\usepackage{booktabs}       
\usepackage{amsfonts}       
\usepackage{nicefrac}       
\usepackage{microtype}      
\usepackage{lipsum}
\usepackage{fancyhdr}       
\usepackage{graphicx}       
\graphicspath{{media/}}     

\usepackage{siunitx}
\usepackage{amssymb}
\usepackage[table,xcdraw]{xcolor}
\usepackage{amsmath}
\usepackage[english]{babel}
\usepackage{subfigure}
\usepackage{paralist} 
\usepackage{hyperref} 
\usepackage{fancyhdr} 
\usepackage{booktabs}
\usepackage{multirow}
\usepackage{amsmath,bm}    
\usepackage{tabularx}      
\usepackage{makecell}      
\newcommand{\best}[1]{\textbf{\boldmath #1}}  
\usepackage{algorithm}
\usepackage{algpseudocode}
\usepackage{pdflscape}

\title{Diffusion Model-Based Posterior Sampling in Full Waveform Inversion}

\author{
  Mohammad H.~Taufik and Tariq~Alkhalifah \\
  Physical Science and Engineering Division \\
  King Abdullah University of Science and Technology (KAUST) \\
  Thuwal 23955, Saudi Arabia \\
  \texttt{mohammad.taufik@kaust.edu.sa, tariq.alkhalifah@kaust.edu.sa}
}

\begin{document}
\maketitle

\begin{abstract}
Bayesian full waveform inversion (FWI) offers uncertainty-aware subsurface models; however, posterior sampling directly on observed seismic shot records is rarely practical at the field scale because each sample requires numerous wave-equation solves. We aim to make such sampling feasible for large surveys while preserving calibration, that is, high uncertainty in less illuminated areas. Our approach couples diffusion-based posterior sampling with simultaneous-source FWI data. At each diffusion noise level, a network predicts a clean velocity model. We then apply a stochastic refinement step in model space using Langevin dynamics under the wave-equation likelihood and reintroduce noise to decouple successive levels before proceeding. Simultaneous-source batches reduce forward and adjoint solves approximately in proportion to the supergather size, while an unconditional diffusion prior trained on velocity patches and volumes helps suppress source-related numerical artefacts. We evaluate the method on three 2D synthetic datasets (SEG/EAGE Overthrust, SEG/EAGE Salt, SEAM Arid), a 2D field line, and a 3D upscaling study. Relative to a particle-based variational baseline, namely Stein variational gradient descent without a learned prior and with single-source (non-simultaneous-source) FWI, our sampler achieves lower model error and better data fit at a substantially reduced computational cost. By aligning encoded-shot likelihoods with diffusion-based sampling and exploiting straightforward parallelization over samples and source batches, the method provides a practical path to calibrated posterior inference on observed shot records that scales to large 2D and 3D problems.
\end{abstract}


\section{Introduction}

Full waveform inversion (FWI) embodies the state-of-the-art (SOTA) framework for seismic velocity model building. This iterative optimization process aims to extract a high-resolution subsurface velocity model by minimizing the discrepancy between observed and simulated data governed by the wave equation \cite{virieux2009overview,tarantola1984inversion}. Yet, the very features that make FWI so compelling also attract practical challenges: severe nonlinearity and cycle skipping when low frequencies are scarce, incomplete illumination, and the high computational burden of repeatedly solving large forward/adjoint problems across many shots. One potential way to temper the cost is to encode or combine shots so that one wavefield evaluation stands in for many; however, the resulting simultaneous-sources crosstalk must be managed throughout the inversion \cite{krebs2009fast,romero2000phase,schiemenz2013accelerated,xue2016seismic}. In practice, this sets up a simple trade-off: encoding cuts cost by performing simultaneous simulations across multiple shots, but the mixing (crosstalk) can leak into the model unless we restrain it. This motivates a Bayesian treatment that quantifies and explores the null space of the solution—where multiple geologically plausible models explain the observed seismic data—through posterior samples and uncertainty maps.

A Bayesian formulation places a prior on the model and defines a likelihood using the wave equation \cite{stuart2010inverse,tarantola2005inverse}; in practice, we access this through the adjoint–state (FWI) gradient of the log–likelihood and thus perform posterior inference via gradient information. Particle transports such as the Stein variational gradient descent (SVGD) algorithms are attractive because they move a set of particles toward the posterior using gradients of the log posterior \cite{liu2016stein}, and they have proven practical for large–scale FWI relative to Markov chain Monte Carlo (MCMC) \cite{zhang2020high,zhang2021introduction,zhang20233}. Many recent SVGD–for–FWI studies adopt uniform (``null'') priors, so the posterior is largely likelihood–dominated; this simplifies implementation but shifts regularization to algorithmic choices (kernel bandwidth, step schedules) and box bounds, which can be mode–seeking and under–estimate posterior variance in ill–posed regimes \cite{izzatullah2024physics,cen2024fwi}. Likelihood annealing—i.e., tempering the data term by a factor $\beta\!\in[0,1]$ and increasing $\beta$ over stages—can stabilize updates, and carefully designed wave–equation perturbations can aid exploration \cite{corrales2025annealed}. However, neither tactic changes the main cost driver: each update still requires full–shot adjoint–state gradients, and with a null prior, the inference remains likelihood–dominated and sensitive to cycle skipping. This motivates a formulation that (i) reduces per–iteration PDE cost via simultaneous (encoded) shots, and (ii) injects a learned prior so uncertainty is governed by both data and geology rather than by algorithmic heuristics alone.

Deep generative priors offer a complementary path. In Plug–and–Play (PnP) \cite{venkatakrishnan2013plug} and Regularization by Denoising (RED) \cite{romano2017little}, a learned denoiser provides a powerful, data–driven regularizer inside an optimization loop. Building on this idea, several works have used unconditional diffusion models as learned priors for deterministic FWI and reported higher–quality velocity reconstructions and better data fits than classical penalties \cite{wang2023prior,taufik2024learned,zhang2024diffusionvel}, complementing conventional regularization theory \cite{menke2018geophysical,aster2018parameter}. Beyond unconditional priors, controllable or conditional variants inject auxiliary information to steer the generated geology \cite{wang2024controllable}; for example, \cite{orozco2024machine} conditions a diffusion generator on common–image–gathers to perform variational inference, while \cite{wang2024wavediffusion} explores a latent diffusion model that maps directly from measured shot gathers to velocity via a shared latent representation. Taken together, these strands mainly deliver either point estimates (regularized optimization) or amortized reconstructions. The latter entails that the generated samples may be plausible, but do not necessarily fit the observed seismic data. To move from regularized reconstruction to explicit posterior sampling with diffusion priors and common-shot gathers data, we turn next to diffusion-based samplers that incorporate measurement information during inference.

To perform posterior sampling with diffusion models as prior, \cite{kawar2022denoising} introduced the Denoising Diffusion Restoration Models (DDRM) for linear inverse problems with closed-form conditioning. For general nonlinear settings, \cite{chung2022diffusion} introduced the Diffusion Posterior Sampling (DPS) by guiding the reverse process with likelihood information. \cite{zhang2025daps} introduced the decoupled annealed posterior sampling (DAPS), extending the DPS framework by decoupling the consecutive steps in the DPS updates with stochastic refinement steps, and showed better posterior sampling exploration and quality. These methods, however, are typically evaluated where the forward operator is cheap or linear (e.g., seismic inversion \cite{ravasi2025geophysical}). In contrast, FWI embeds an expensive, nonlinear relationship between the model and data, which is computationally more demanding by an order of magnitude than the reverse diffusion step. Therefore, to do posterior sampling in FWI, two obstacles remain: (i) the physics likelihood is expensive (requires a large number of PDE solves); and (ii) deterministic or weakly stochastic guidance can under-estimate posterior variance if treated as pure optimization rather than a Markov transition.

Our perspective is to couple diffusion priors with wave-equation likelihoods in a way that is both computationally viable and statistically calibrated for large-scale applications. We build on a decoupled annealing view of diffusion sampling: at each noise level, we (i) predict a clean model with the diffusion network, (ii) perform stochastic clean-space refinement using the physical likelihood, and (iii) reintroduce noise to move to the next level. Decoupling the refinement from the reverse diffusion update enables large, nonlocal corrections under strong nonlinearity and preserves stochastic mixing between levels, both of which are difficult when guidance is tightly coupled to small reverse steps. At the same time, we exploit simultaneous-sources (encoded-shot) data with unbiased likelihood scaling to reduce forward/adjoint solves by approximately the number of shots in the supergather, while the diffusion prior helps suppress the crosstalk introduced by encoding. We will interchangeably use supergather and encoded-shot data to refer to the same seismic data with multiple source locations.

We develop and evaluate a diffusion–based posterior sampling framework tailored to FWI. Specifically, our contributions from this work include: 

1. \textbf{A decoupled, encoded-shot diffusion sampler for FWI.} We design a DAPS-style posterior sampler that (i) predicts a clean model at each diffusion level, (ii) performs stochastic clean-space Langevin refinement using encoded (simultaneous-sources) data, and (iii) renoises to the next level with Denoising Diffusion Probabilistic Model (DDPM) \cite{ho2020denoising} variance. Decoupling preserves stochastic mixing and allows non-local corrections under strong nonlinearity, while encoded shots reduce forward/adjoint solves by $\approx m$ and the diffusion prior helps suppress source-mixing crosstalk \cite{krebs2009fast,romero2000phase}.\\
2. \textbf{Scalable evaluation in 2D/3D and on field data.} We train unconditional diffusion priors in 2D (patches) and 3D (cubes) and assess the sampler on three 2D synthetics, a 2D field line, and a 3D upscaling study. We report both velocity model-space metrics and data-space metrics, together with computational cost analysis (PDE solves and diffusion network forward evaluations).\\
3. \textbf{Head-to-head comparison with a variational baseline.} Using SVGD as a strong conventional VI baseline utilizing the same initial models and schedules, we compare accuracy, computational cost, and posterior statistics. We clarify that, in our sampler, the inner stochastic refinement move is the Markov kernel, whereas SVGD’s exploration is governed mainly by the Stein kernel and particle interactions \cite{liu2016stein}. Ablations show that deterministic guidance (e.g., DDIM with zero variance) underestimated variance, while stochastic re-noise (DDPM variance) maintains posterior fidelity \cite{ho2020denoising,song2020score}.

We begin by explaining the theory behind the proposed methodology. We then begin the empirical analysis by detailing the diffusion prior training (2D/3D). We then present 2D synthetic, 2D field, and 3D upscaling results, including posterior diagnostics and computational cost analysis. We conclude by highlighting the current properties and limitations of our framework and discussing potential extensions.

\section{Methodology}

This section establishes the forward modeling setup and data notation, then promotes the Bayesian viewpoint adopted for inversion through diffusion prior. We first introduce the simultaneous-source (encoded) strategy used to reduce wave-equation cost, outline the diffusion prior and its reverse parameterization for generating plausible velocity models, and present a decoupled inference procedure that alternates clean-space refinement with reverse diffusion. We end the section with a brief discussion of computational cost and practical scheduling choices.

\subsection{Forward model, data, and likelihood}
Let $x \in \mathbb{R}^{n}$ denote the subsurface model to be inferred (here, we focus on the acoustic velocity field). For a given source index $i \in \{1,\dots, N_s\}$, the seismic modeling operator $F_i(\cdot)$ maps $x$ to a predicted shot gather $F_i(x)$ by numerically solving (in our case) the acoustic wave equation on the acquisition geometry used in the survey. Let $d_i$ be the observed shot gather for source $i$. For simplicity, we restrict the following Bayesian formulation of FWI to our 2D synthetic data examples, which utilizes a Gaussian assumption and a simple mean-squared-error objective function.

In other words, we assume additive, zero-mean measurement noise with variance $\sigma_y^2$ per sample and define the full-shot Gaussian negative log-likelihood (data misfit) as
\begin{equation}
\Phi(x) \;=\; \frac{1}{2\sigma_y^2}\,\sum_{i=1}^{N_s} \big\| d_i - F_i(x) \big\|_2^2 ,
\label{eq:phi-full}
\end{equation}
where $\|\cdot\|_2$ denotes the Euclidean norm after stacking time and receiver samples.

\paragraph{Bayesian formulation of FWI.}
In this setting, we view the problem as Bayesian inference on the model $x$ given the observed data $\{d_i\}_{i=1}^{N_s}$. 
With the Gaussian likelihood in \eqref{eq:phi-full}, the data term is
\begin{equation}
p(d \mid x) \;\propto\; \exp\!\big(-\Phi(x)\big).
\label{eq:likelihood}
\end{equation}
A prior over plausible geology, $p(x)$, is represented implicitly by an unconditionally trained denoising diffusion model. 
The posterior then reads
\begin{equation}
p(x \mid d) \;\propto\; \exp\!\big(-\Phi(x)\big)\, p(x).
\label{eq:posterior}
\end{equation}
Our sampler targets draw from \eqref{eq:posterior} by alternating (i) diffusion reverse steps that respect $p(x)$ and (ii) short clean-space refinements that reduce $\Phi(x)$ using computationally efficient encoded shots.

\paragraph{Encoded (simultaneous-sources) shots.}
At each guided diffusion level, we form an encoded-shots data (supergather) by drawing a mini-batch $B \subset \{1,\dots, N_s\}$ with $|B|=m$, where $m$ denotes the total number of supergathers, and random encoding weights (e.g., polarity flip) $w_i \sim \mathcal{N}(0,1)$, with
\begin{equation}
\tilde d \;=\; \sum_{i \in B} w_i\, d_i,
\qquad
\tilde F(x) \;=\; \sum_{i \in B} w_i\, F_i(x).
\label{eq:encoded-defs}
\end{equation}
The encoded-shots misfit is
\begin{equation}
\tilde\Phi(x) \;=\; \frac{1}{2\sigma_y^2}\, \big\| \tilde d - \tilde F(x) \big\|_2^2.
\label{eq:phi-encoded}
\end{equation}
Its gradient, computed by the adjoint-state method with one forward and one adjoint solve for the encoded wavefield, yields an unbiased estimator of the full-shot gradient:
\begin{equation}
\widehat{\nabla \Phi}(x) \;=\; \frac{N_s}{m}\,\nabla \tilde\Phi(x),
\label{eq:unbiased-grad}
\end{equation}
since $\mathbb{E}_{w,B}\!\left[\frac{N_s}{m}\,\nabla \tilde\Phi(x)\right] \!=\! \nabla \Phi(x)$ when the $w_i$ have zero-mean, unit-variance and the batch $B$ is uniformly sampled.

\subsection{Diffusion prior and reverse parameterization}
We represent prior information on plausible geology with an unconditionally trained denoising diffusion model using the DDPM sampler. Let $\{\beta_t\}_{t=1}^T$ be a variance schedule with $\alpha_t = \prod_{s=1}^{t} (1-\beta_s)$. The forward diffusion corrupts a clean model $x_0$ into $x_t$; the reverse process uses a learned network $g_\theta(x_t,t)$ that predicts the clean model, i.e., we adopt the ``predict-$x_0$'' parameterization
\begin{equation}
\hat x_0 \;=\; g_\theta(x_t,t),
\label{eq:x0-pred}
\end{equation}
with inputs and outputs normalized to $[-1,1]$ during training. For inference, we map models to physical units before taking likelihood gradients and re-normalize when stepping the reverse process. In line with the previous studies for velocity generation with diffusion models, we start the diffusion inference from the last few timesteps to (i) reduce the number of function evaluations and (ii) utilize the kinematically correct initial velocity model instead of starting from random noise.

\subsection{Decoupled diffusion inference with encoded shots data}
At diffusion level $t$ (from $T$ down to $1$), we decouple measurement guidance from the reverse step and insert a short, stochastic clean-space refinement under the FWI likelihood. This includes the following three components:

\paragraph{(i) Predict a clean model.} From the current noisy state $x_t$, obtain $\hat x_0$ via \eqref{eq:x0-pred} and map it to physical units.

\paragraph{(ii) Clean-space Langevin refinement.} Starting at $z^{(0)}=\hat x_0$, take $K_t$ unadjusted Langevin steps using the unbiased encoded-shots gradient \eqref{eq:unbiased-grad}:
\begin{equation}
z^{(k+1)} \;=\; z^{(k)} \;-\; \eta_t\, \widehat{\nabla \Phi}\!\left(z^{(k)}\right) \;+\; \sqrt{2\,\eta_t}\; \xi^{(k)},
\end{equation}
and
\begin{equation}
\xi^{(k)} \sim \mathcal{N}(0,I),
\quad
k=0,\dots,K_t-1.
\label{eq:langevin}
\end{equation}
Here $\eta_t>0$ is a guidance step size (in physical units), $m$ is the number of encoded sources in \eqref{eq:encoded-defs}, and $\sigma_y^2$ is as in \eqref{eq:phi-full}. Denote the refined clean model by $z^{(K_t)}$.

\paragraph{(iii) Re-noise to the next level.} Return to the diffusion space and draw
\begin{equation}
x_{t-1} \;\sim\; \mathcal{N}\!\big(\tilde \mu_t(x_t, z^{(K_t)}),\; \tilde \beta_t I \big),
\label{eq:renoise}
\end{equation}
where the DDPM posterior parameters are
\begin{equation}
\tilde \beta_t \;=\; \frac{1-\alpha_{t-1}}{1-\alpha_t}\,\beta_t,
\label{eq:ddpm-posterior0}
\end{equation}
and
\begin{equation}
\tilde \mu_t(x_t, x_0) \;=\; 
\frac{\sqrt{\alpha_{t-1}}\,\beta_t}{1-\alpha_t}\, x_0 \;+\;
\frac{\sqrt{1-\beta_t}\,(1-\alpha_{t-1})}{1-\alpha_t}\, x_t.
\label{eq:ddpm-posterior}
\end{equation}

With these three, the proposed diffusion model-based posterior sampling algorithm can be summarized in Table \ref{alg:proposed}.

\begin{table}[t]
\caption{The proposed framework}
\label{alg:proposed}
\centering
\begin{tabularx}{\columnwidth}{@{}X@{}}
\toprule
\begin{minipage}{\columnwidth}
\begin{algorithmic}[1]
  \State \textbf{Input:} Forward operator $F$, observed data $d$, model $\epsilon_\theta$, schedule $\{\alpha_t\}$, Langevin steps $K$
  \State \textbf{Output:} Sample $x_0 \sim p(x_0 \mid d)$
  \State Initialize $x_T \sim \mathcal{N}(0, I)$
  \For{$t = T$ to $1$}
    \State $\hat{x}_0 = \dfrac{1}{\sqrt{\alpha_t}}\Bigl(x_t - \sqrt{1 - \alpha_t}\,\epsilon_\theta(x_t, t)\Bigr)$ \Comment{Predict denoised sample}
    \For{$k = 1$ to $K$} \Comment{Stochastic refinement steps}
      \State $\hat{x}_0 \leftarrow \hat{x}_0 + \eta \nabla_{\hat{x}_0} \lVert d - F(\hat{x}_0)\rVert^2 + \nu\,\mathcal{N}(0, I)$
    \EndFor
    \If{\texttt{renoise}}
      \State $x_{t-1} \sim q(x_{t-1} \mid \hat{x}_0)$ \Comment{Re-noise refined sample}
    \EndIf
  \EndFor
  \State \textbf{return} $x_0$
\end{algorithmic}
\end{minipage}\\
\bottomrule
\end{tabularx}
\end{table}

Compared to the original DAPS framework, we modify the stochastic refinement step such that the noise and data likelihood updates have independent noise, and we make the renoising step optional (i.e., treating it as an additional hyperparameter). The motivation behind the first modification is to account for the amplitude discrepancy between the likelihood and noise terms. Note the standard gradient-based optimization can be used as the stochastic refinement step, yielding a deterministic updates in which the posterior exploration stage only concentrates around the maximum a posteriori solution. 

\subsection{Computational cost}
Each encoded-shots gradient in \eqref{eq:langevin} requires one forward and one adjoint solve (for the encoded wavefield). If $\mathcal{T}$ denotes the set of guided levels, the total number of PDE solves per posterior sample is
\begin{equation}
\text{PDE solves} \;\approx\; 2\, m \sum_{t \in \mathcal{T}} K_t .
\label{eq:pde-count}
\end{equation}
Compared to full-shot guidance ($m=N_s$), encoded shots reduce cost approximately in proportion to the supergather size $m$, while the diffusion prior mitigates source-mixing crosstalk introduced by encoding.

\section{Numerical experiments}

In the following, we begin by sharing the training details of the 2D and 3D diffusion models. In all of the diffusion training, we consider training these models from scratch and work solely in the velocity model domain. The examples are arranged to highlight three main features. In the first part, we demonstrate the performance of the proposed framework when dealing with 2D synthetic data to highlight the quality of the results. We then consider a towed-stream field data application to understand how it handles unknown measurement noise and its performance as a function of different diffusion model training distributions. We end the examples by further highlighting the scalability of our framework using 3D synthetic data with a varying acquisition area. We end this section by assessing the quality of the produced samples in the 2D examples, analyzing whether these samples do indeed explain the observed seismic data.

\subsection{Diffusion model training}

\begin{figure}
  \centering
  \subfigure[]{
    \includegraphics[width=\columnwidth]{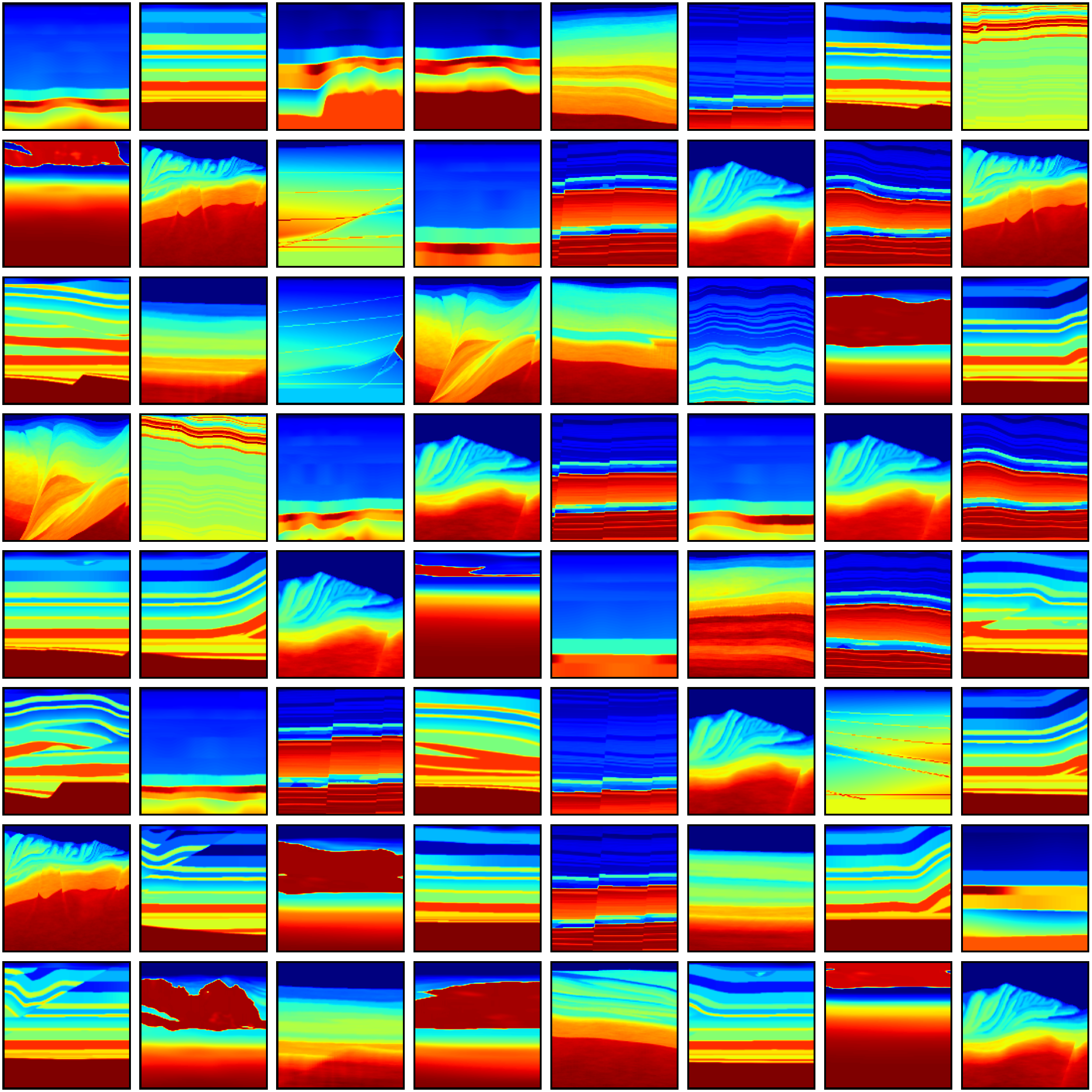}}
  \caption{Samples from the training (\textsc{Realistic2D}) dataset for the 2D diffusion model.}
  \label{data_2d}
\end{figure}

We train two diffusion models on $128\times128$ velocity patches using the denoising diffusion probabilistic model (DDPM) formulation configured with the $v$-prediction objective \cite{salimans2022progressive}. The first model utilized 30{,}311 training patches drawn from SEG Open Data velocity models and several in-house models (hereafter \textsc{Realistic2D}) shown in Figure \ref{data_2d}. The second diffusion model utilizes a suite of procedurally generated random models that contain mostly layered media, whose trends are constructed such that they possess a velocity range between 1.5 and 4.5 km/s (hereafter \textsc{Random2D}). Velocities are normalized to $[-1,1]$ using per-sample minimum and maximum values. The denoiser is a 2D U-Net with base width 128, stage multipliers $\{1,2,4,8,16\}$, and eight residual blocks. We train with 1000 diffusion steps (linear schedule) for 52k iterations, batch size $1$ with gradient accumulation of $2$ (effective batch $\approx2$), Adam optimizer at $5\times10^{-6}$, and an $\ell_1$ loss on $v$.

\begin{figure}
  \centering
  \subfigure[]{
    \includegraphics[width=\columnwidth]{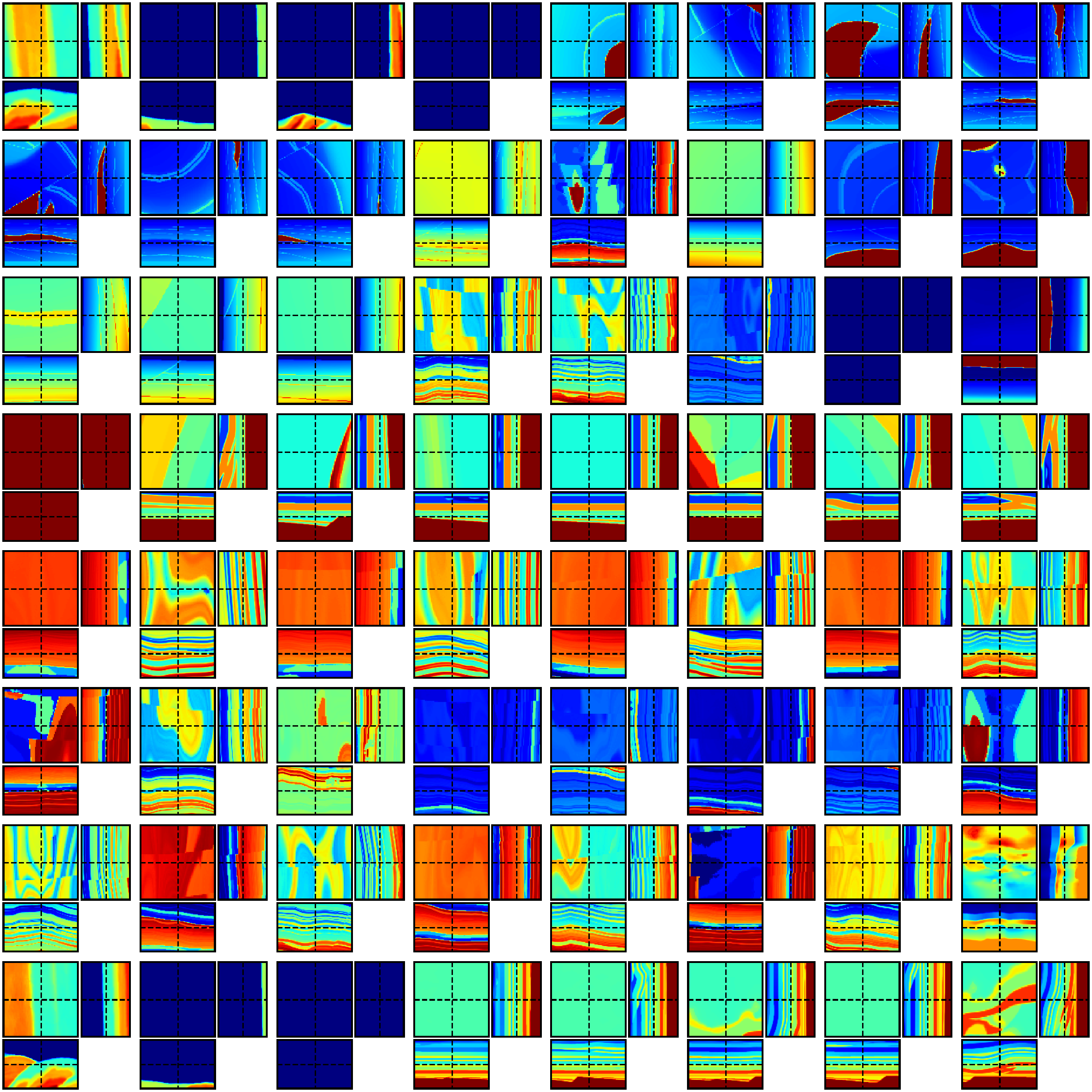}}
  \caption{Samples from the training (\textsc{Realistic3D}) dataset for the 3D diffusion model.}
  \label{data_3d}
\end{figure}

We train a 3D pixel-space diffusion prior on cubic subvolumes of size $64^3$ extracted from the same family of velocity models used in the \textsc{Realistic2D} set shown in Figure \ref{data_3d}. The denoiser is a 3D U-Net with base width 128 channels (stage multipliers as in 2D), unlike the 2D diffusion, the 3D U-Net is predicting the noise level $\epsilon$ and schedule as in 2D (1000 steps, linear noise schedule, Adam with an $\ell_1$ loss; learning rate $1\times10^{-6}$; batch size $2$ with gradient accumulation of $2$ for an effective batch of $\approx4$). Standard 3D augmentations (axis flips and $90^\circ$ rotations) are applied, and velocities are normalized to $[-1, 1]$ using the same strategy as in the 2D diffusion model.

Both the 2D and 3D diffusion models training and inference are executed on a machine with an Intel Xeon Gold 6230R (52 cores), 252 GB RAM, and a single A100 80 GB. In all synthetic and field tests that follow, we evaluate out-of-distribution behavior: none of the evaluation velocity models appear in the training corpus. Throughout the following examples, we consider only training the three diffusion models once and apply them to different observed data.

\subsection{Posterior sampling using 2D ocean-bottom node synthetic data}

Equipped with the trained 2D diffusion model, we first conduct three synthetic data experiments in a variety of geological conditions. In all of the following three examples, we start the reverse diffusion process from the respective (normalized) initial velocity model from the last 300 timesteps. To accommodate the velocity shape mismatch between the diffusion model and a velocity model of arbitrary size, we utilize a patchwise sampling strategy. Specifically, we simply extract overlapping patches with a stride of 128. We inject 5 FWI iterations as guidance every 5 reverse diffusion timesteps. The FWI optimization utilizes the Adam optimizer with a learning rate of 50. We utilize 20 different velocity model realizations (particles) to compute the mean and standard deviation, which conventionally would cost 20 FWI applications.

\begin{figure}
  \centering
  \subfigure[Initial velocity\label{arid_init}]{
    \includegraphics[width=0.4\columnwidth]{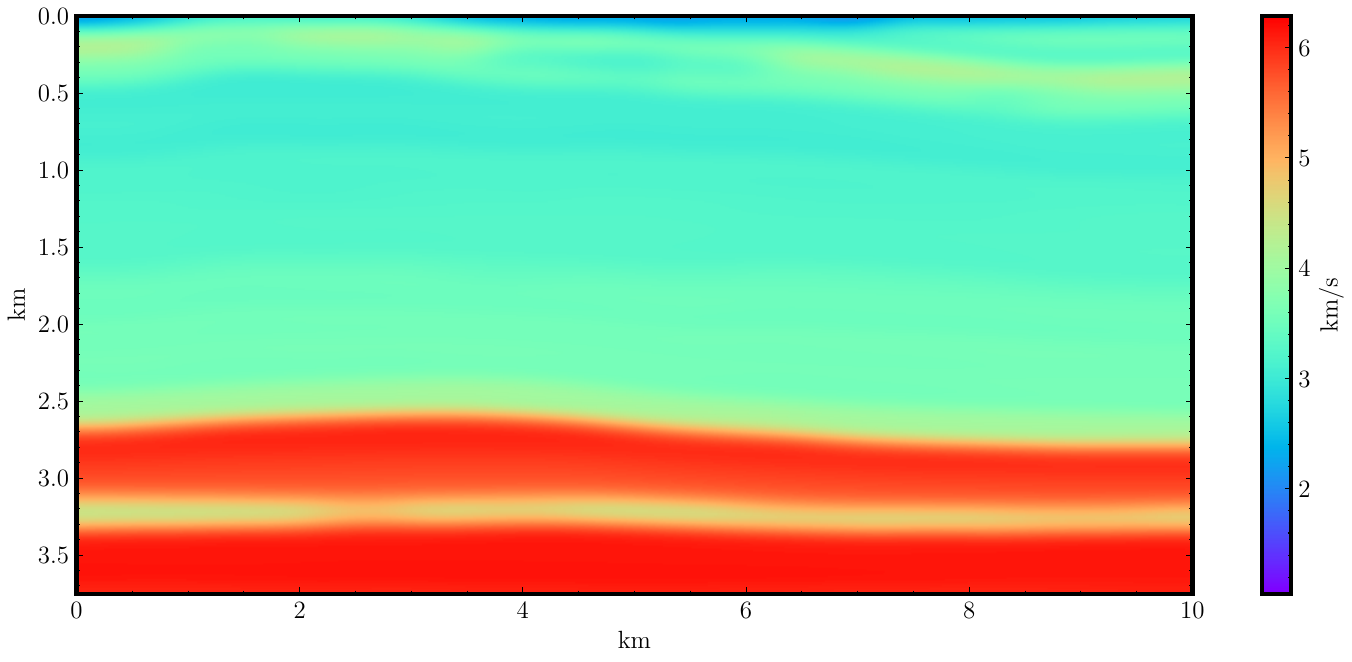}}
  \subfigure[Ground truth\label{arid_true}]{
    \includegraphics[width=0.4\columnwidth]{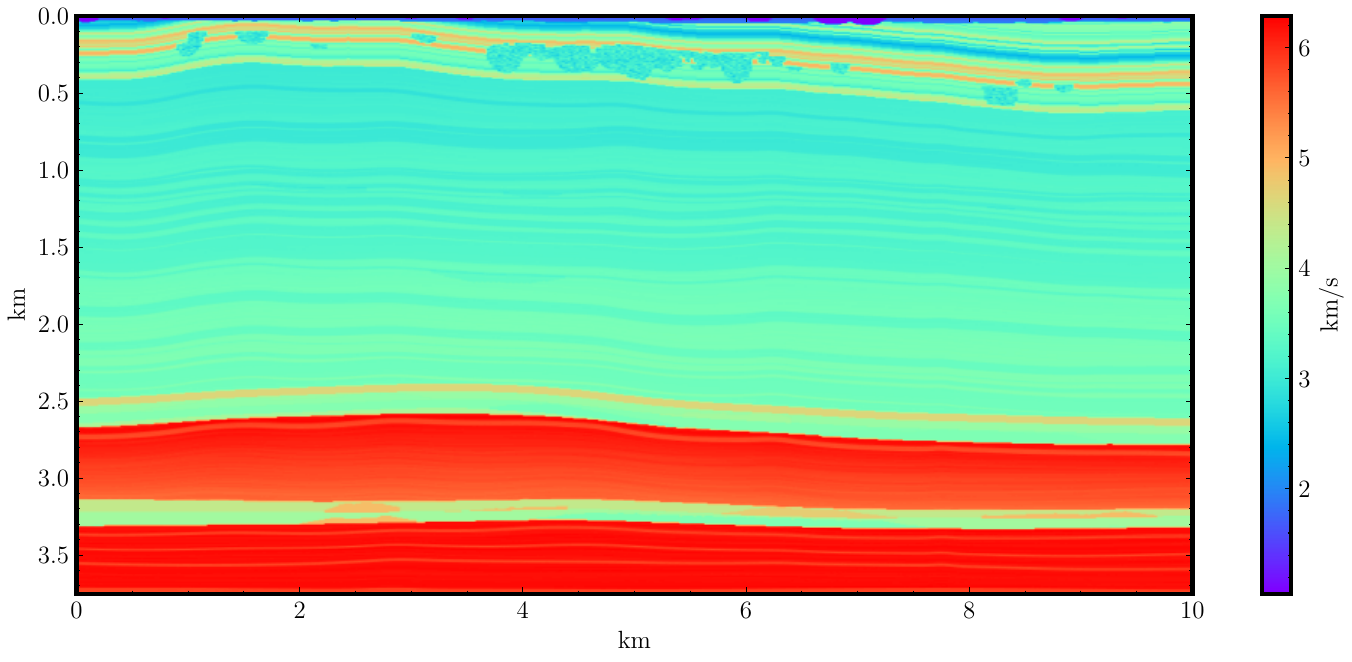}}

  \subfigure[Diffusion posterior mean \label{arid_mean}]{
    \includegraphics[width=0.4\columnwidth]{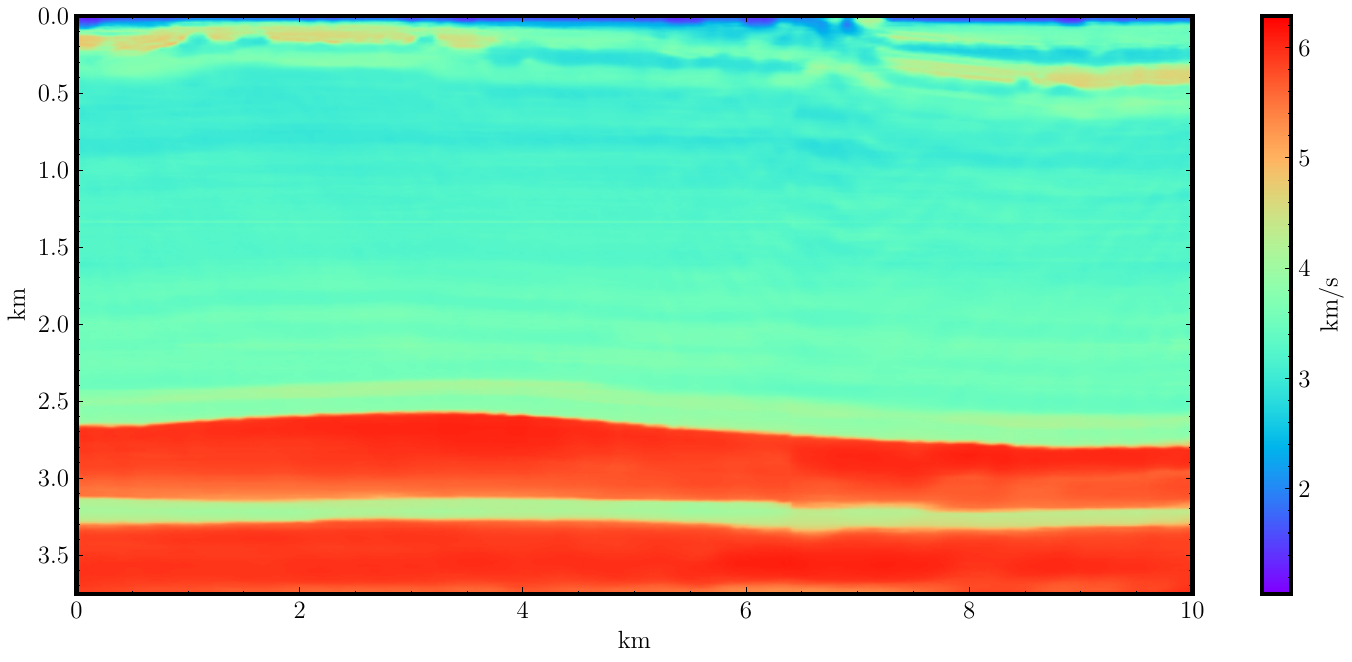}}
  \subfigure[SVGD mean (conventional FWI gradients)\label{arid_mean_svgd}]{
    \includegraphics[width=0.4\columnwidth]{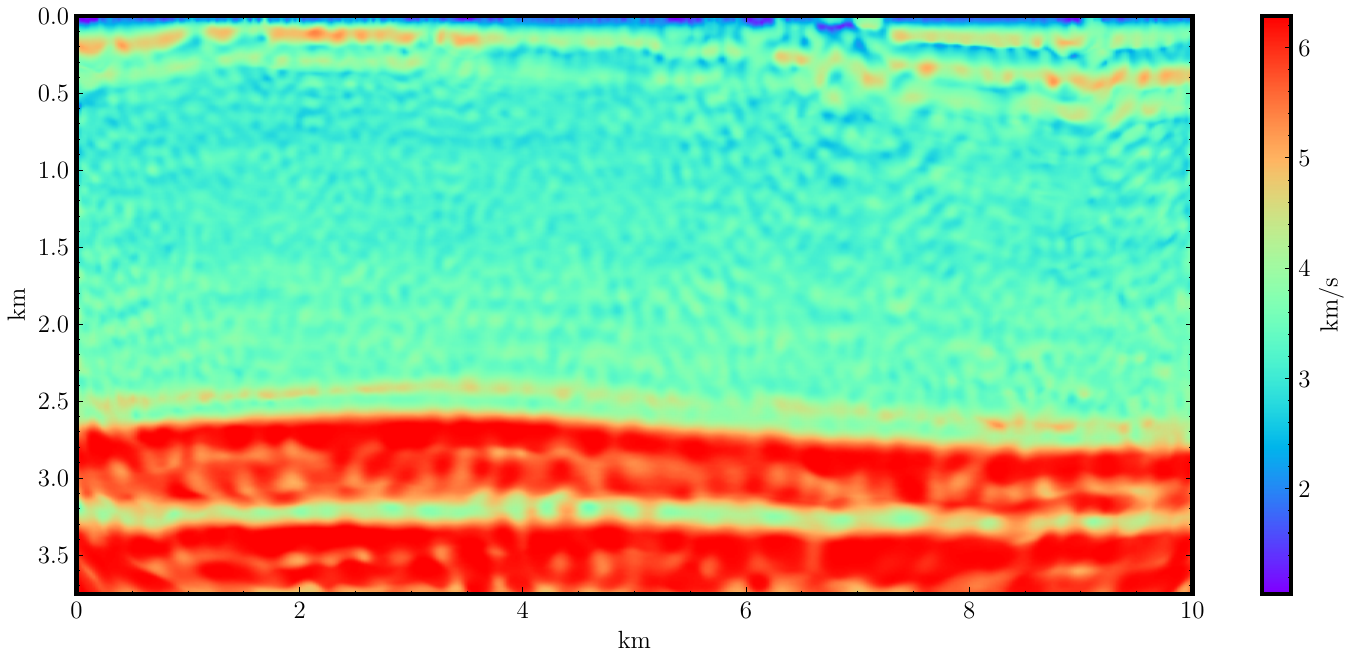}}

  \subfigure[Diffusion posterior standard deviation\label{arid_std}]{
    \includegraphics[width=0.4\columnwidth]{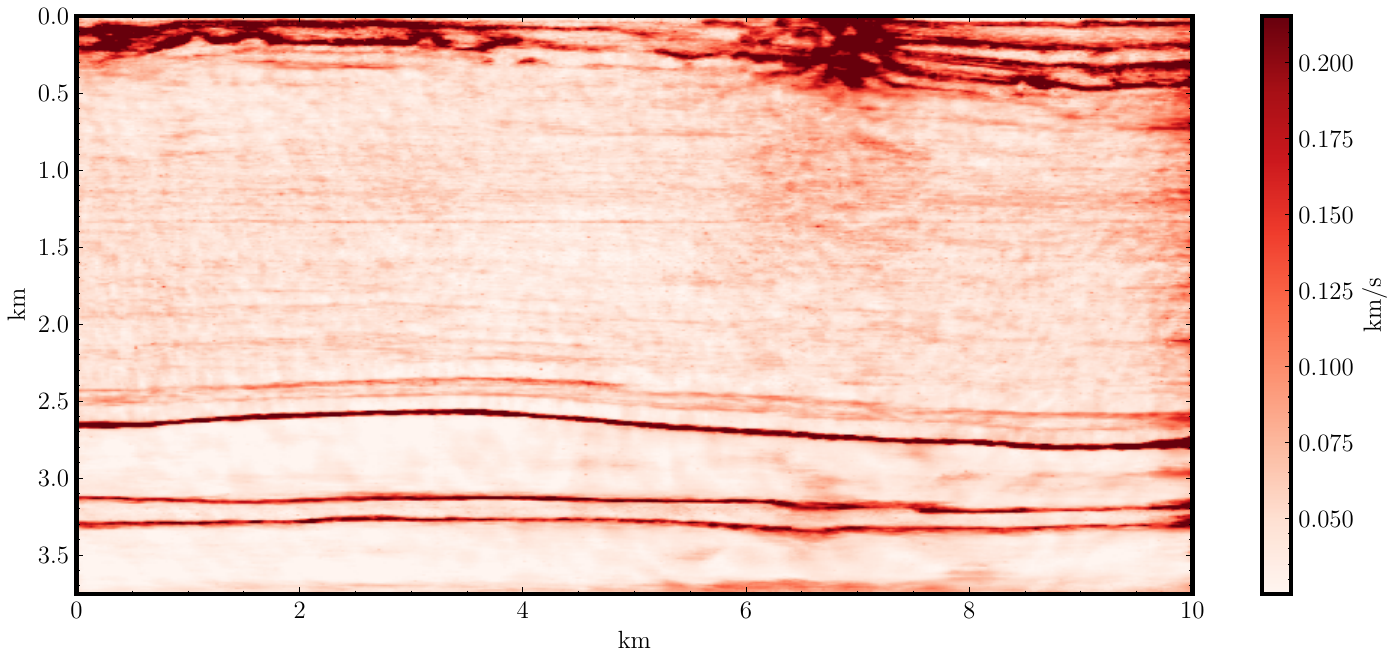}}
  \subfigure[SVGD posterior standard deviation\label{arid_std_svgd}]{
    \includegraphics[width=0.4\columnwidth]{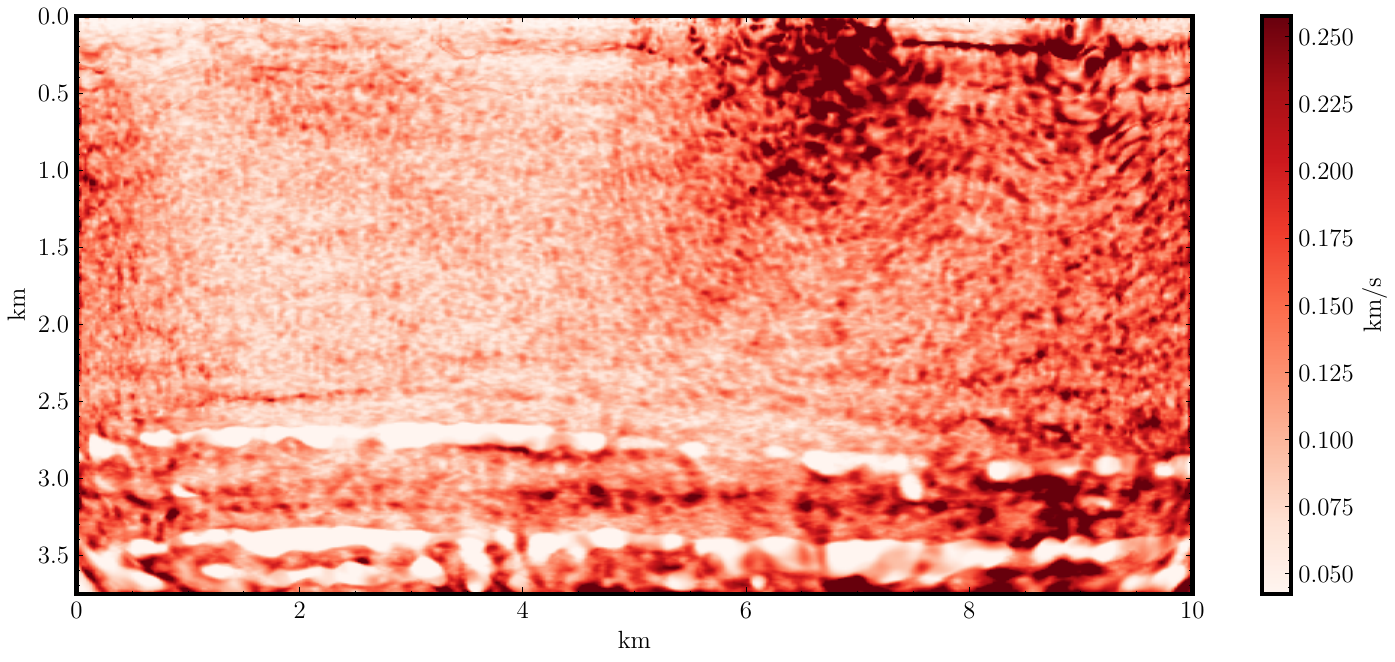}}

  \caption{SEAM Arid synthetic (2D OBN). (a) Initial model; (b) ground truth; (c) diffusion posterior mean with multi-source FWI guidance; (d) SVGD mean with conventional FWI gradients; (e–f) corresponding posterior standard deviations, respectively.}
  \label{arid_2d}
\end{figure}

In the first experiment, we study the performance of our framework in which the adjoint-state FWI gradient is contaminated by strong numerical artefacts. To do so, we utilize the SEAM Arid model, which contains a very low-velocity layer in its first few hundred meters mimicking a typical karst layer in an arid area (Figure \ref{arid_2d}). The 2D velocity model is of size 400 $\times$ 600 with a lateral and vertical grid spacing of 25 and 6.25 m, respectively. The 600 nodes are placed near the surface at a depth of 25 m with a regular spacing of 25 m. The 128 sources are located at the same depth as the receiver nodes with a grid spacing of $\approx$117 m. To perform the simultaneous-source FWI, we form 4 supergathers, each containing randomly selected source locations whose selection is updated every FWI iteration. To generate the synthetic data, we utilize a Ricker wavelet of 6 Hz and perturb the simulated data with Gaussian noise. We perform 300 simultaneous-source FWI iterations using a single frequency band with an initial velocity model obtained by smoothing the true velocity model with a Gaussian filter (Figure \ref{arid_2d}).

As shown in Figure \ref{arid_2d}, the proposed framework provides a more representative sample of the posterior distribution compared to the SVGD algorithm with the same 20 particles and standard non-simultaneous-source FWI data. Specifically, by comparing the standard deviation maps (Figures \ref{arid_2d}e and \ref{arid_2d}f), we can clearly observe the influence of poor FWI gradient (data likelihood), and the limited number of particles, on the performance of SVGD. In this case, the presence of the karst layer as well as the complex near-surface lithology of the model hinders SVGD from converging to a good posterior estimate. In contrast, courtesy of the learned prior of the diffusion model, the proposed framework manages to overcome these challenges, resulting in a much cleaner posterior mean (Figure \ref{arid_2d}c) and much improved structural uncertainty estimates.

\begin{figure}
  \centering
  \subfigure[Initial velocity\label{salt_init}]{
    \includegraphics[width=0.4\columnwidth]{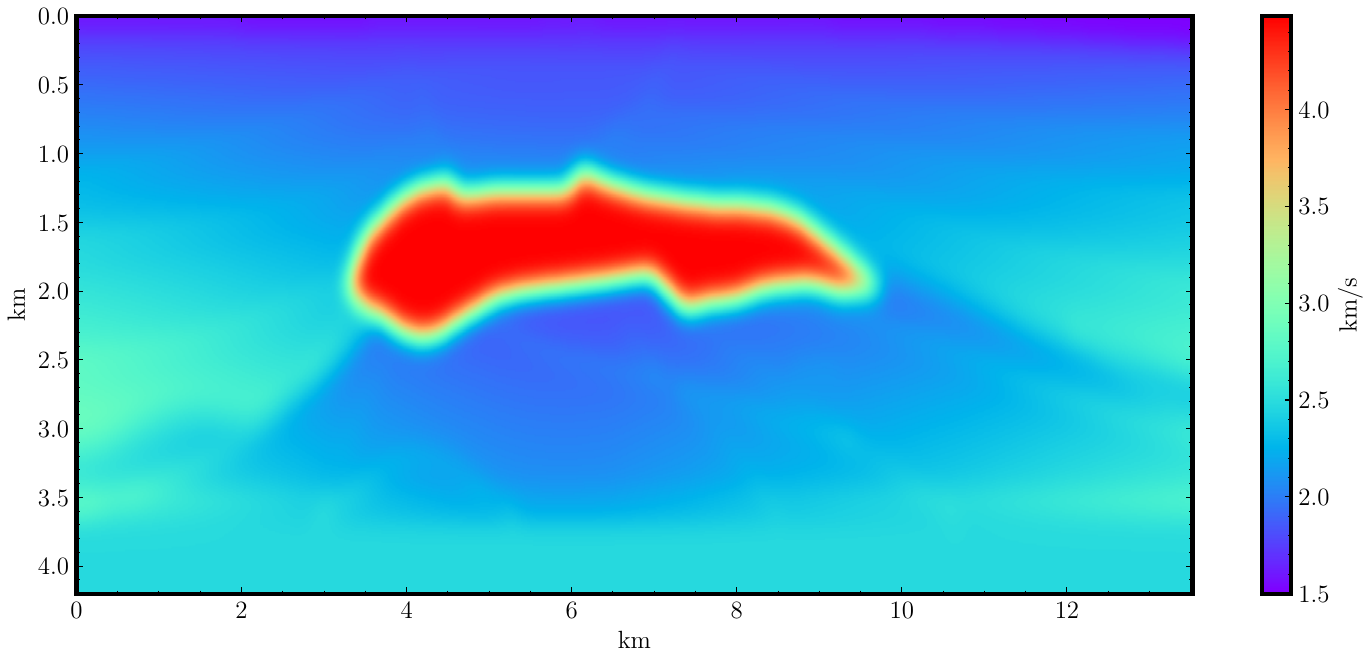}}
  \subfigure[Ground truth\label{salt_true}]{
    \includegraphics[width=0.4\columnwidth]{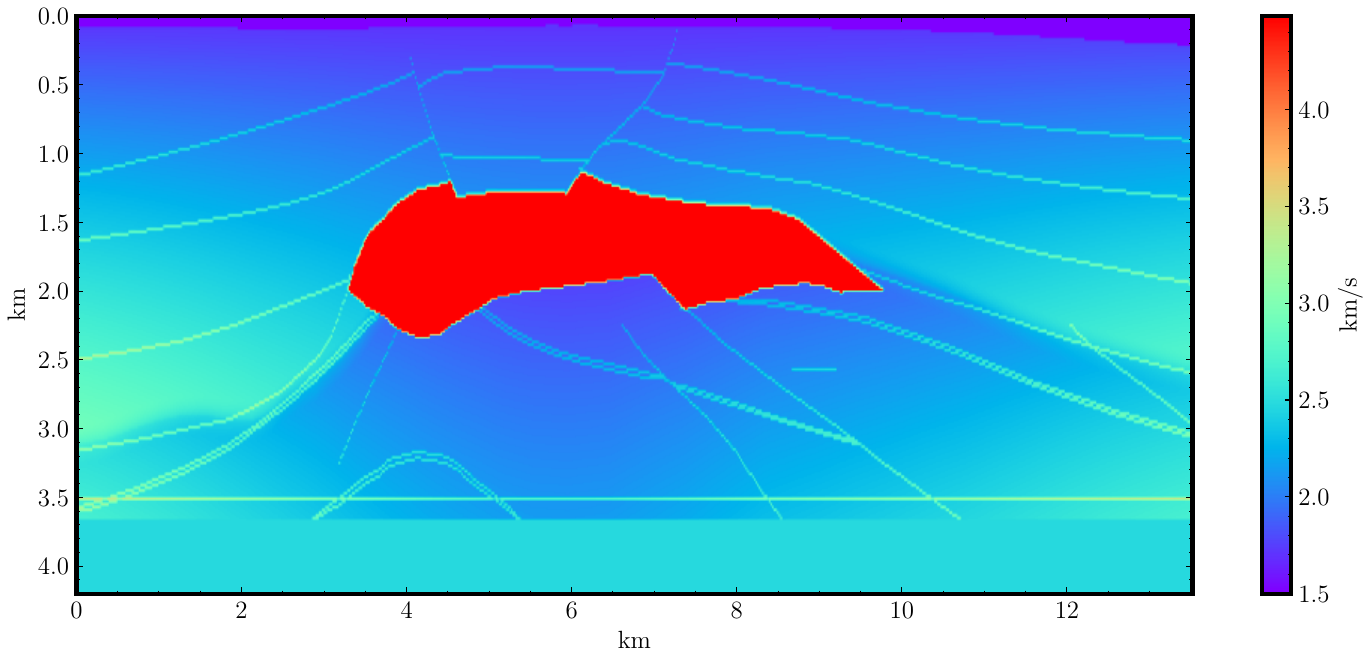}}

  \subfigure[Diffusion posterior mean \label{salt_mean}]{
    \includegraphics[width=0.4\columnwidth]{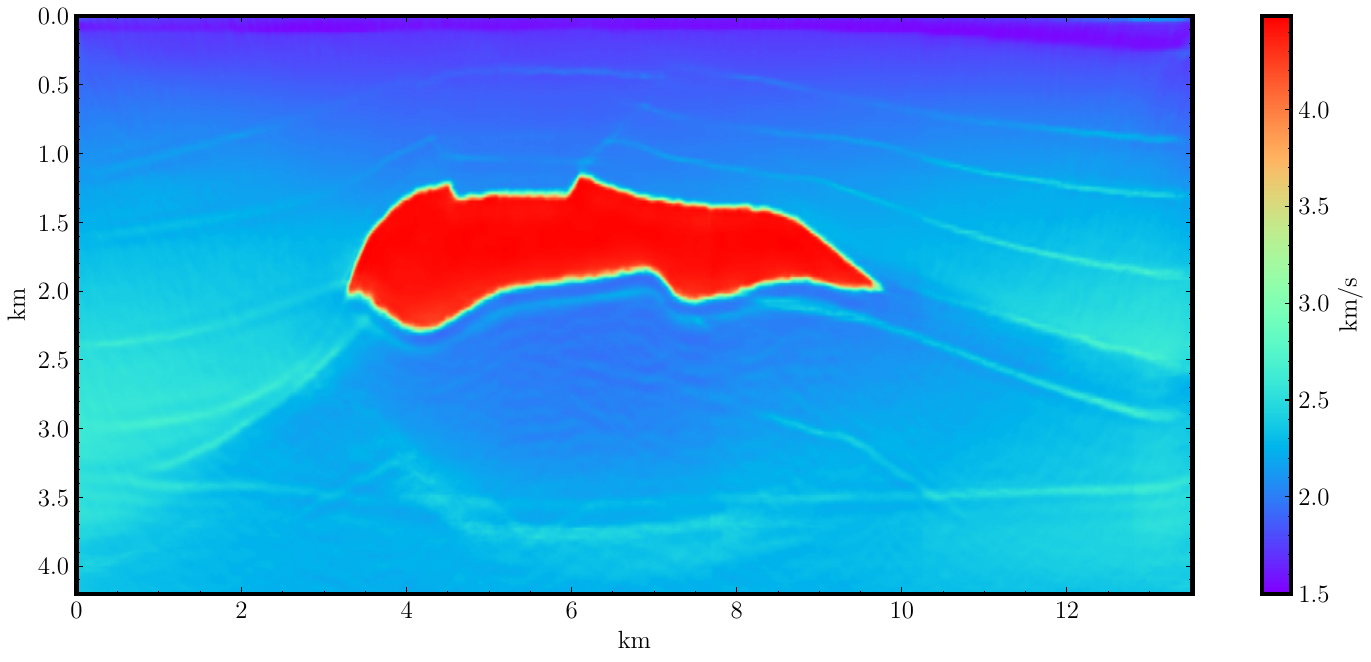}}
  \subfigure[SVGD mean (conventional FWI gradients)\label{salt_mean_svgd}]{
    \includegraphics[width=0.4\columnwidth]{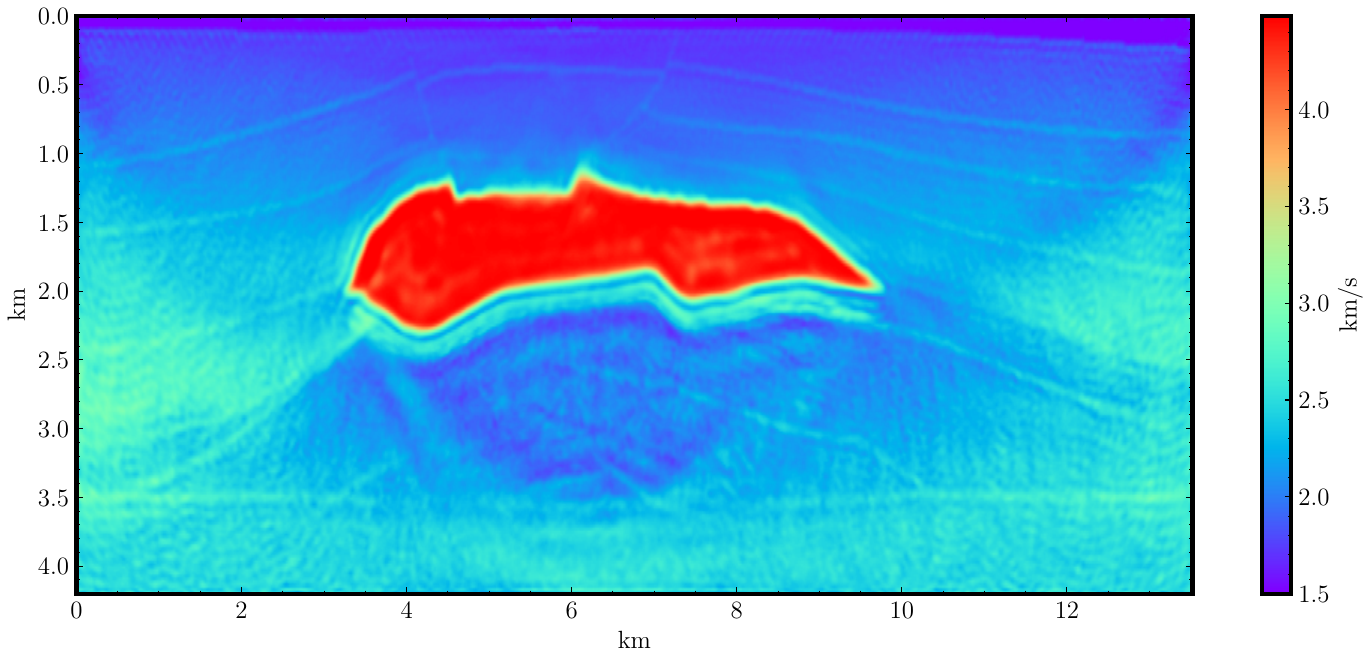}}

  \subfigure[Diffusion posterior standard deviation\label{salt_std}]{
    \includegraphics[width=0.4\columnwidth]{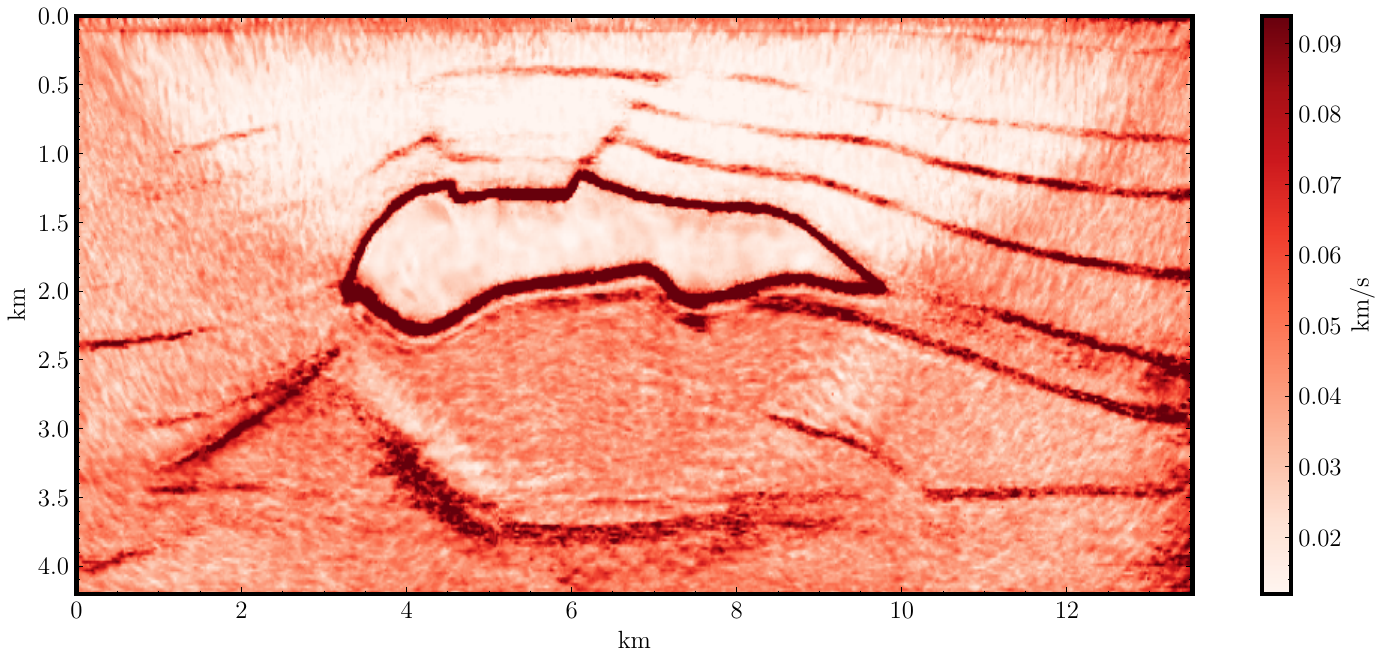}}
  \subfigure[SVGD posterior standard deviation\label{salt_std_svgd}]{
    \includegraphics[width=0.4\columnwidth]{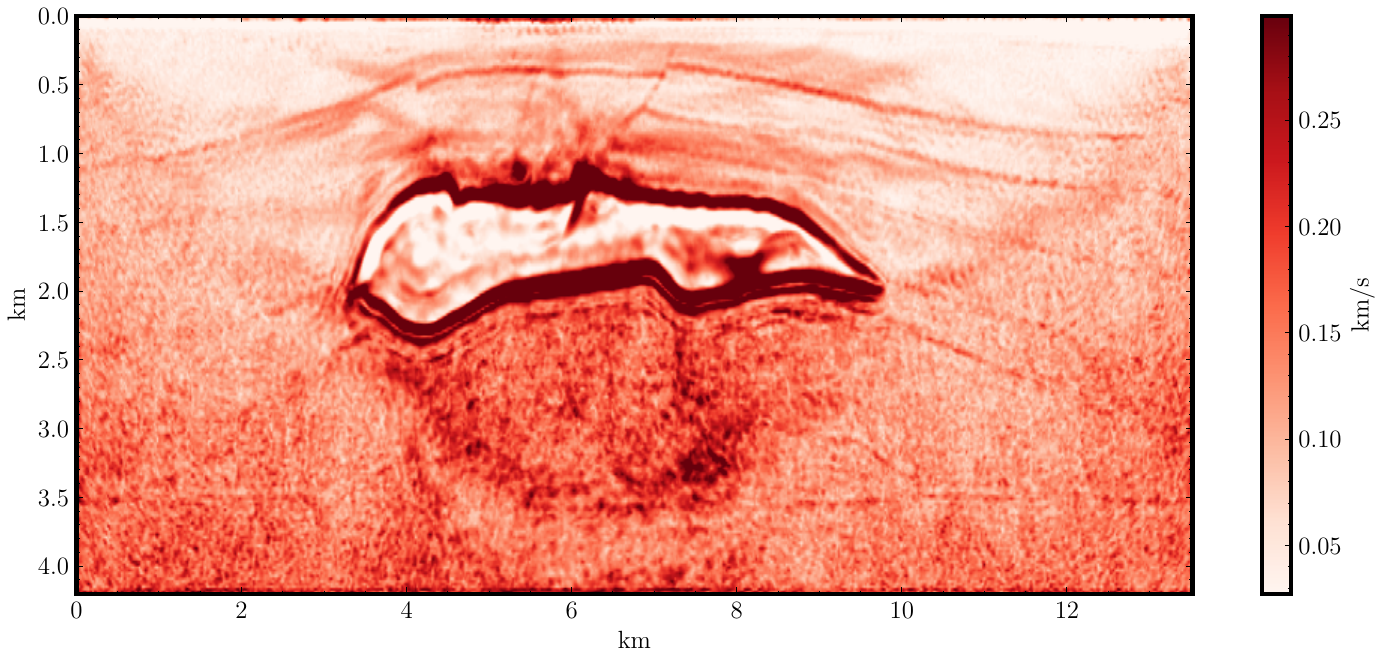}}

  \caption{SEG/EAGE Salt synthetic (2D OBN). (a) Initial model; (b) ground truth; (c) diffusion posterior mean with multi-source FWI guidance; (d) SVGD mean with conventional FWI gradients; (e–f) corresponding posterior standard deviations, respectively.}
  \label{salt_2d}
\end{figure}

We further study the effect of the FWI gradient in a salt diapir environment represented by the SEG/EAGE Salt model (Figure \ref{salt_2d}). The 2D velocity model is of size 210 $\times$ 676 with a lateral and vertical grid spacing of 20 m. The 676 nodes are placed near the surface at a depth of 20 m with a regular spacing of 20 m. The 128 sources are located at the same depth as the receiver nodes with a grid spacing of $\approx$105 m. We utilize the same data frequency, number of supergathers as in the previous case, number of posterior samples, and FWI iterations to perform the inference.

As shown in Figure \ref{salt_2d}, we observe the same phenomenon as in the previous example in that the proposed framework provides more representative samples when compared with the SVGD algorithm. Not only do we manage to suppress the noisy gradient updates coming from strong reverberations of the top salt, but our estimated mean also manages to capture the small-scale feature of the velocity model compared to SVGD. Moreover, the standard deviation map of our framework indicates that we preserve the acquisition-related uncertainty in that high velocity variations are present in areas where the data illuminations are weak.

The last 2D synthetic experiment involves the use of the SEG/EAGE Overthrust model. The 2D velocity model is of size 187 $\times$ 801 with a lateral and vertical grid spacing of 20 m. The 801 nodes are placed near the surface at a depth of 20 m with a regular spacing of 20 m. The 256 sources are located at the same depth as the receiver nodes with a grid spacing of $\approx$78 m. We utilize the same data frequency, number of supergathers as in the previous case, number of posterior samples, and FWI iterations to perform the inference.

\begin{figure}
  \centering
  \subfigure[Initial velocity\label{overthrust_init}]{
    \includegraphics[width=0.4\columnwidth]{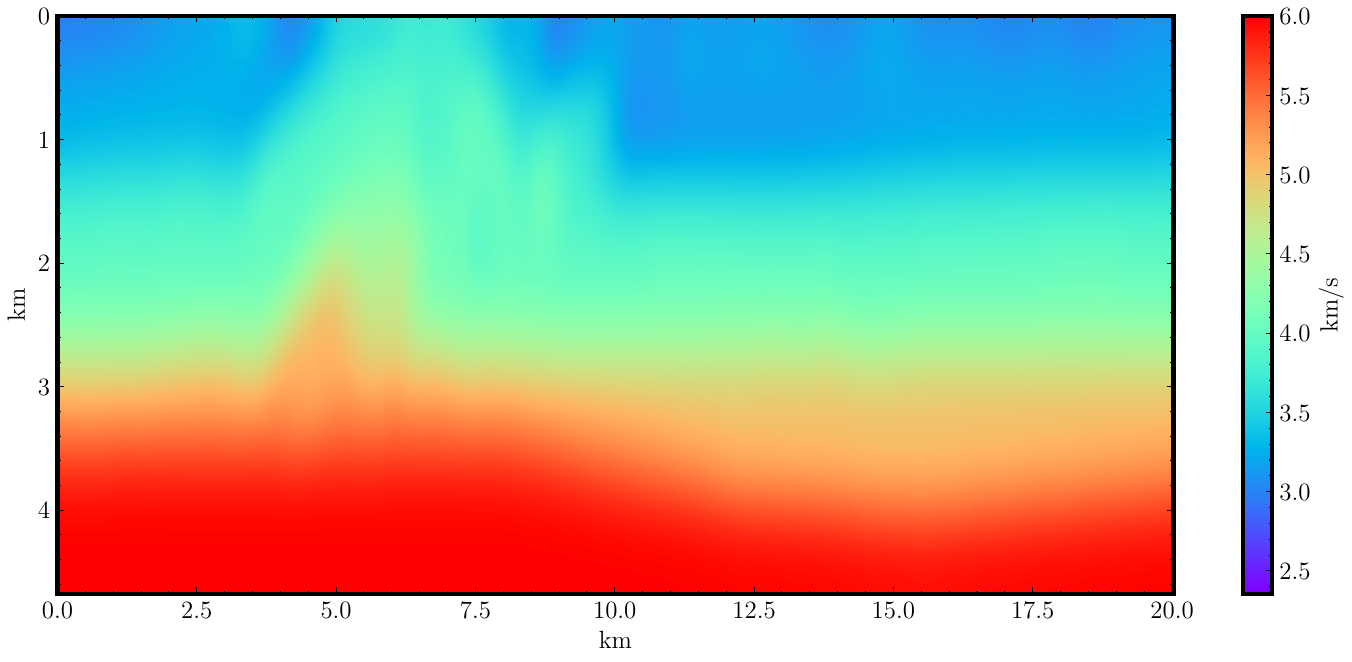}}
  \subfigure[Ground truth\label{overthrust_true}]{
    \includegraphics[width=0.4\columnwidth]{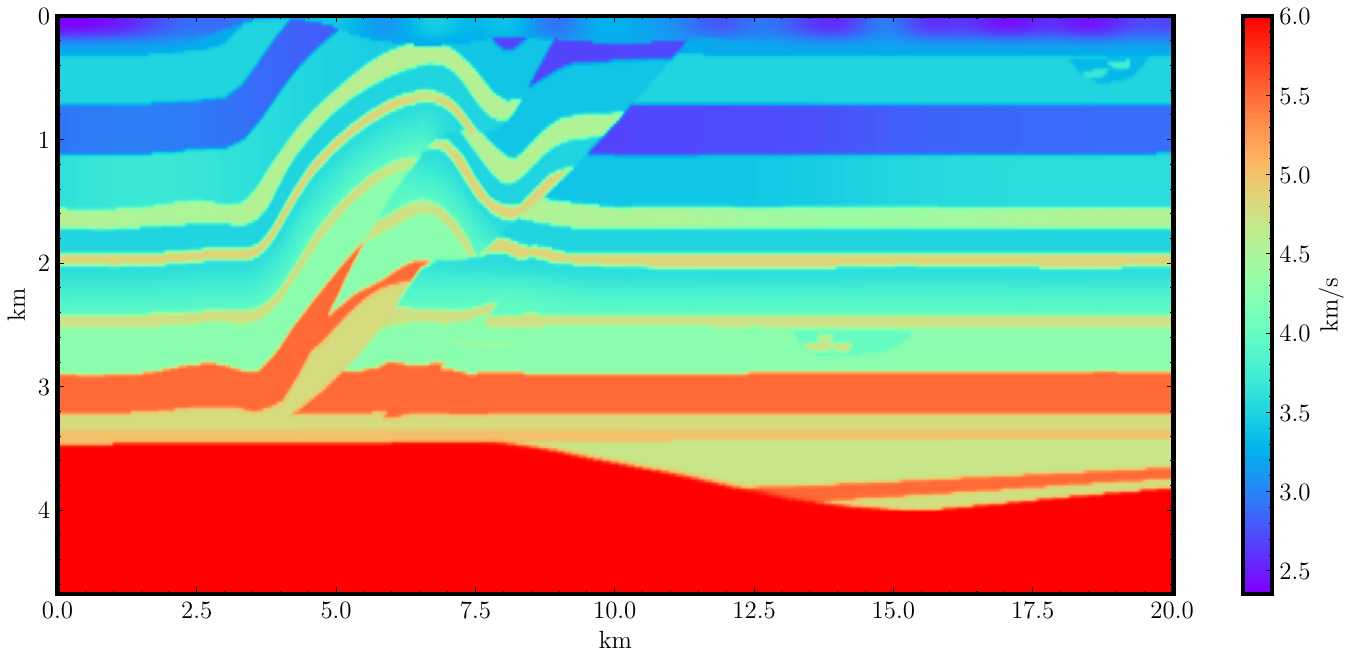}}

  \subfigure[Diffusion posterior mean \label{overthrust_mean}]{
    \includegraphics[width=0.4\columnwidth]{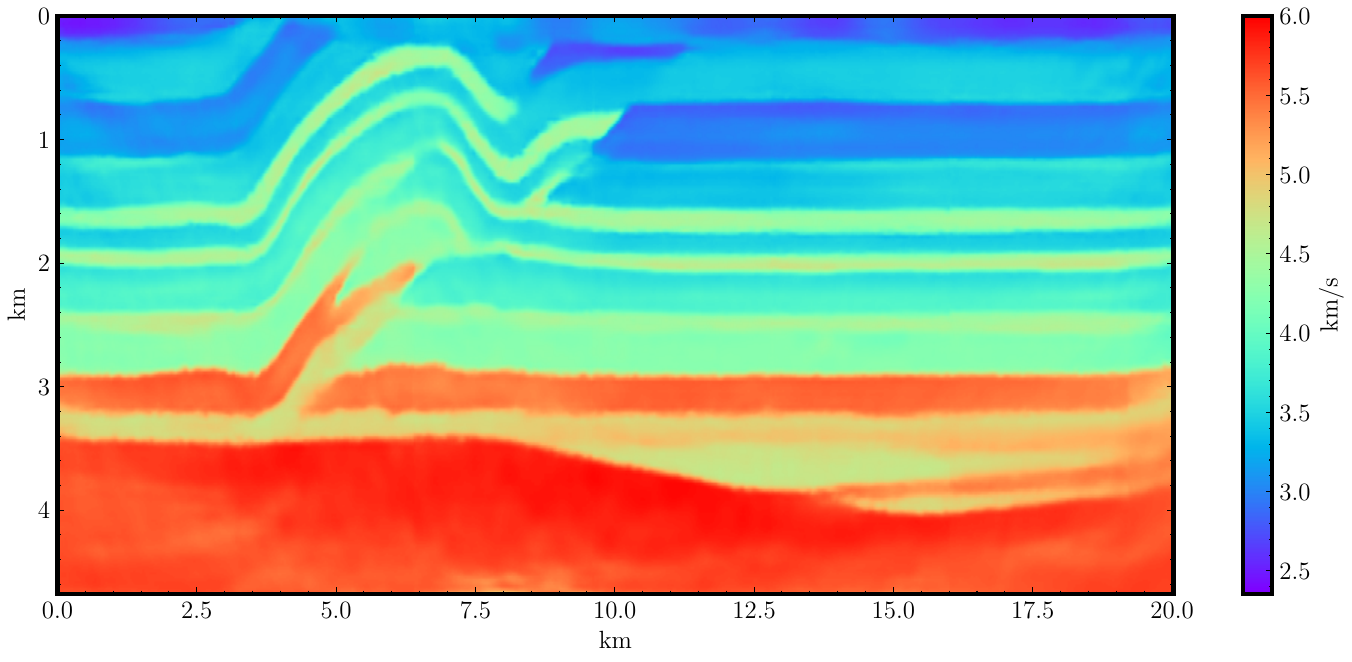}}
  \subfigure[SVGD mean (conventional FWI gradients)\label{overthrust_mean_svgd}]{
    \includegraphics[width=0.4\columnwidth]{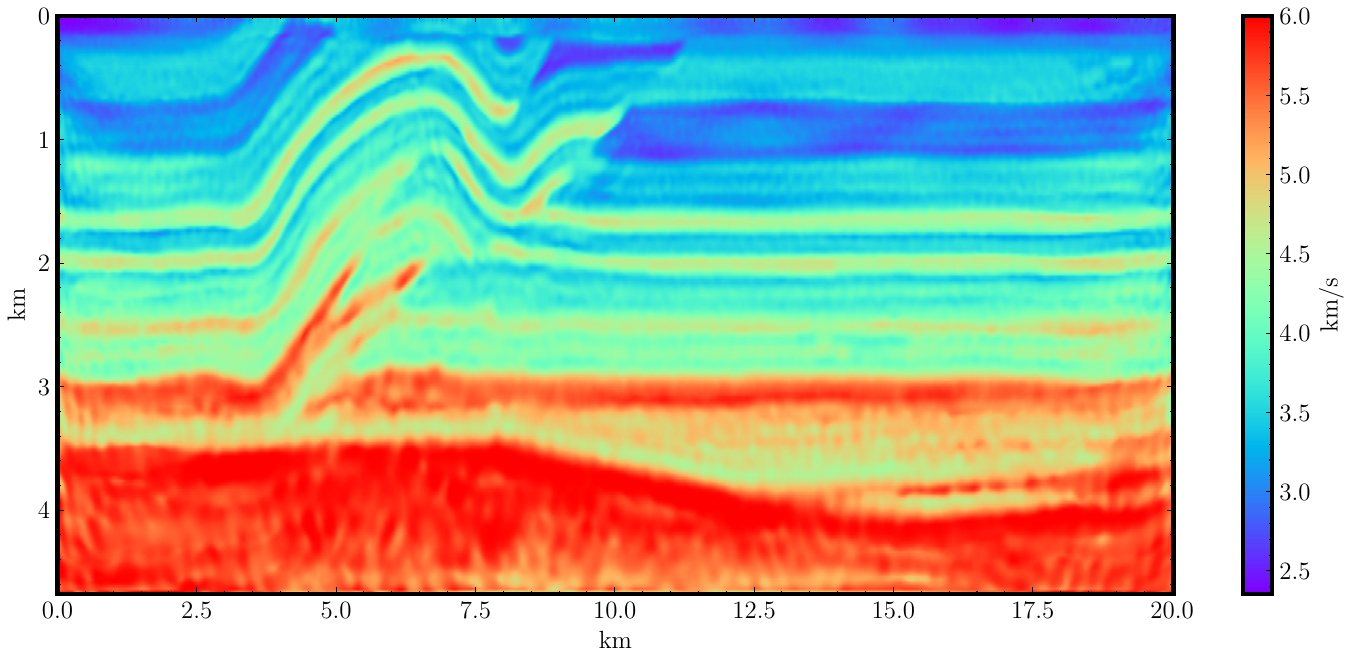}}

  \subfigure[Diffusion posterior standard deviation\label{overthrust_std}]{
    \includegraphics[width=0.4\columnwidth]{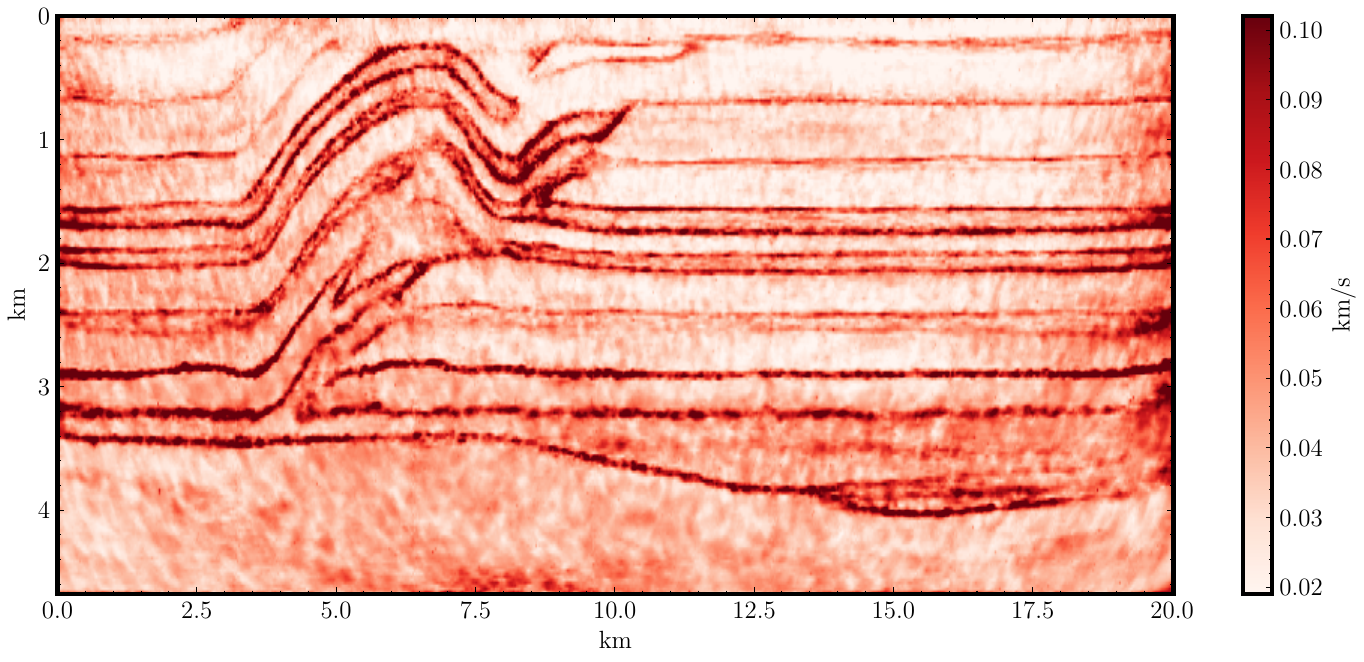}}
  \subfigure[SVGD posterior standard deviation\label{overthrust_std_svgd}]{
    \includegraphics[width=0.4\columnwidth]{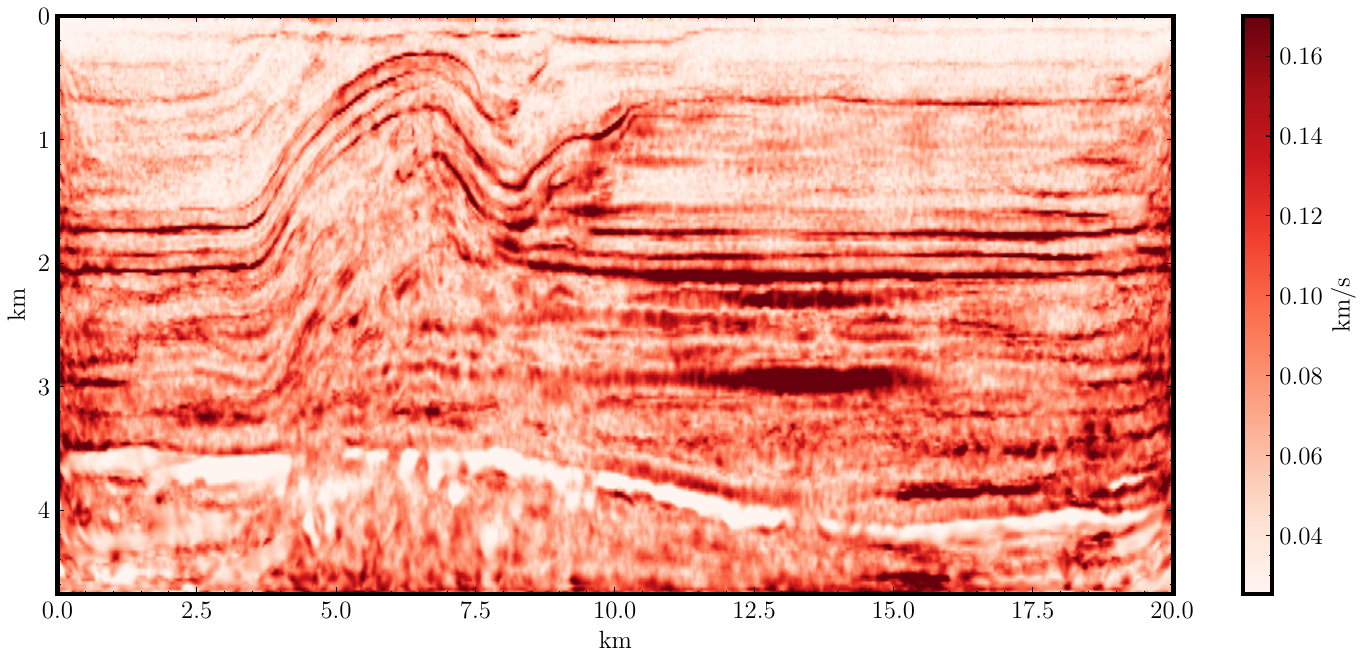}}

  \caption{SEG/EAGE Overthrust synthetic (2D OBN). (a) Initial model; (b) ground truth; (c) diffusion posterior mean with multi-source FWI guidance; (d) SVGD mean with conventional FWI gradients; (e–f) corresponding posterior standard deviations, respectively.}
  \label{overthrust_2d}
\end{figure}

In this case, we aim to study the capability of our framework in a situation when the FWI gradient is well-behaved. Compared to the previous two cases, the difference between our framework and SVGD becomes less pronounced, though the proposed framework still delivers less noisy mean estimates and a better structural uncertainty from its standard deviation map (Figure \ref{overthrust_2d}). This indicates that even when the FWI gradient is well-behaved, the proposed framework delivers a higher perceptual quality than SVGD. These improved results are achieved with a considerable reduction of cost thanks to the simultaneous-sources FWI implementation.

\subsection{Posterior sampling using 2D towed-streamer field data}

To further highlight the performance of our framework in handling unknown measurement noise, we utilize a 2D towed-streamer field dataset from North West Australia acquired by CGG (now Viridien). We utilize the first 576 shot gathers, 648 receivers, with a group interval of approximately 0.0125  km. We perform a bandpass filtering such that the peak frequency is around 6 Hz. For demonstration, we consider the use of a single-band standard FWI process. We follow the same procedure in obtaining the source wavelet, data preconditioning, initial velocity model, and objective function as described in \cite{kalita2017exa}. When performing the diffusion model inference, we start the reverse diffusion process from the respective (normalized) initial velocity model from the last 100 timesteps. To accommodate the velocity shape mismatch between the diffusion model and the desired velocity model size for this area, we simply extract overlapping patches with a stride of 64. We inject 2 FWI iterations as guidance every diffusion timestep, resulting in a total of 200 FWI iterations. The FWI optimization utilizes the Adam optimizer with a learning rate of 5.

\begin{figure}
  \centering
  \subfigure[Initial velocity\label{field_init}]{
    \includegraphics[width=0.4\columnwidth]{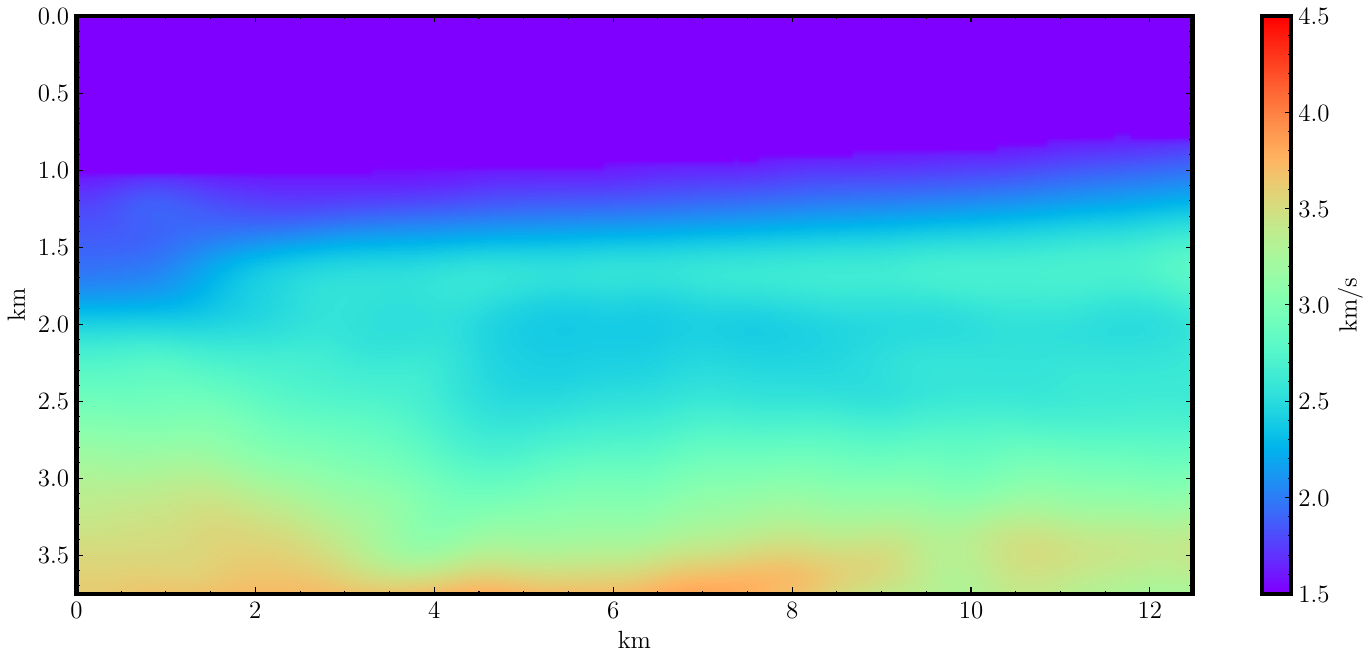}}
  \subfigure[Posterior sample (diffusion)\label{field_sample_random}]{
    \includegraphics[width=0.4\columnwidth]{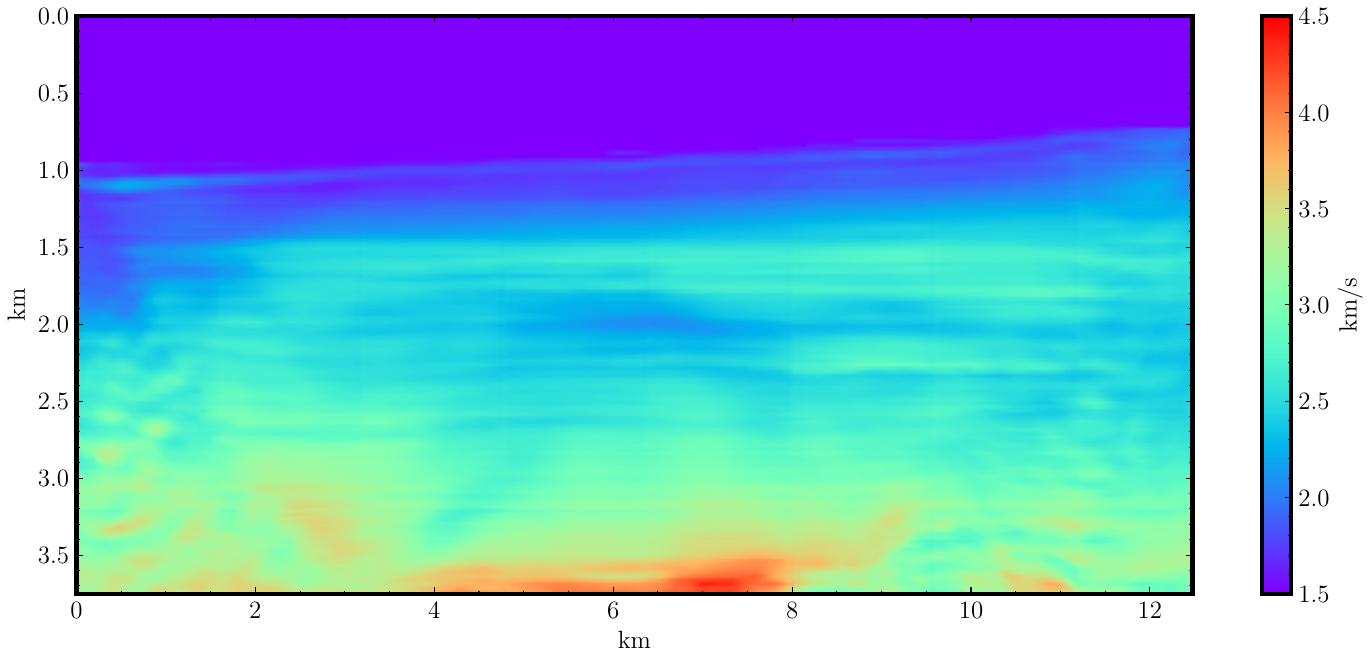}}

  \subfigure[Posterior mean \label{field_mean_random}]{
    \includegraphics[width=0.4\columnwidth]{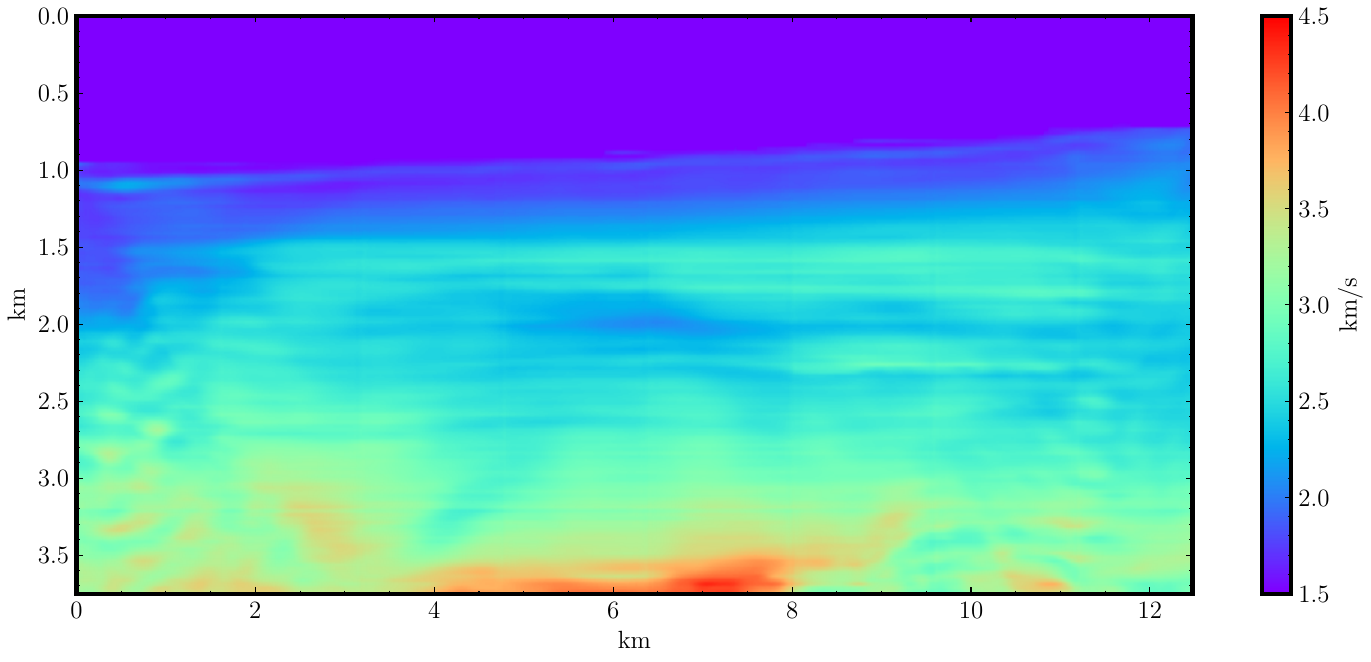}}
  \subfigure[Posterior standard deviation (diffusion)\label{field_std_random}]{
    \includegraphics[width=0.4\columnwidth]{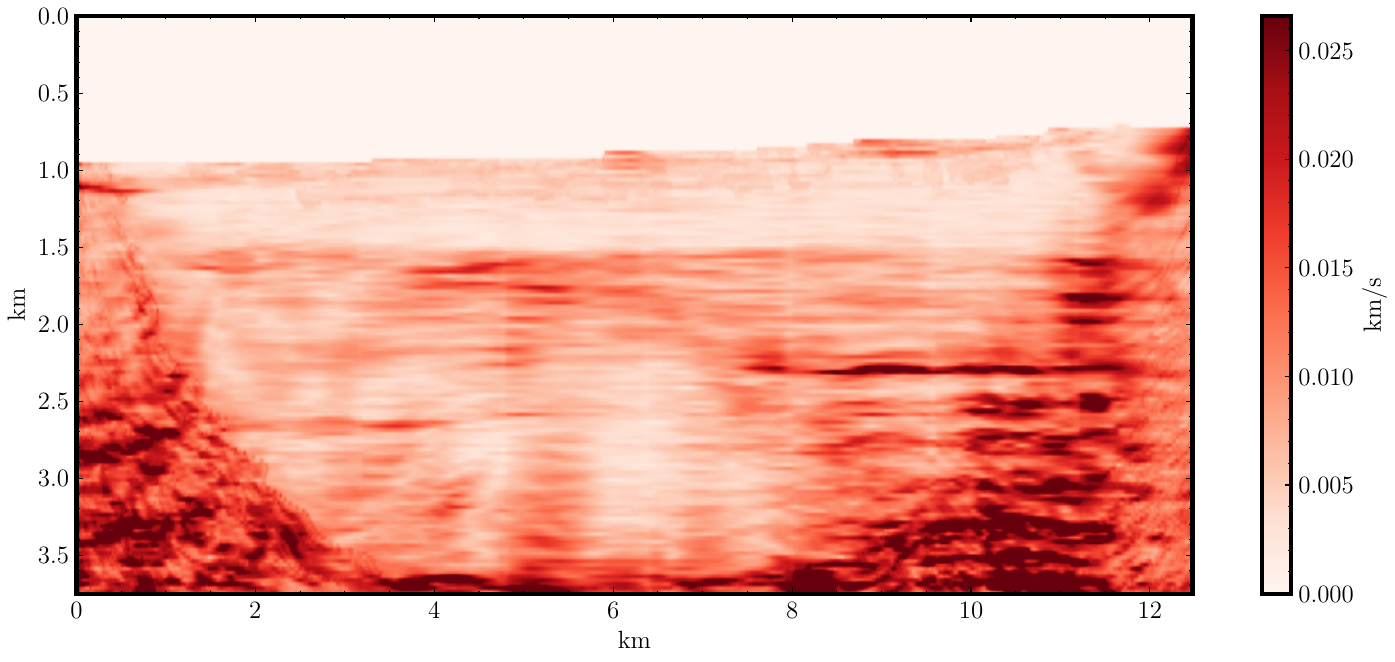}}

  \caption{2D towed-streamer field data with a diffusion prior trained on the \textsc{Random2D} velocity family and simultaneous-source likelihood guidance. (a) Initial model; (b) one posterior sample; (c) posterior mean; (d) posterior standard deviation.}
  \label{field_2d_random}
\end{figure}

\begin{figure}
  \centering
  \subfigure[Initial velocity\label{field_init_seg}]{
    \includegraphics[width=0.4\columnwidth]{Figures/field2d_init.pdf}}
  \subfigure[Posterior sample (diffusion)\label{field_sample_seg}]{
    \includegraphics[width=0.4\columnwidth]{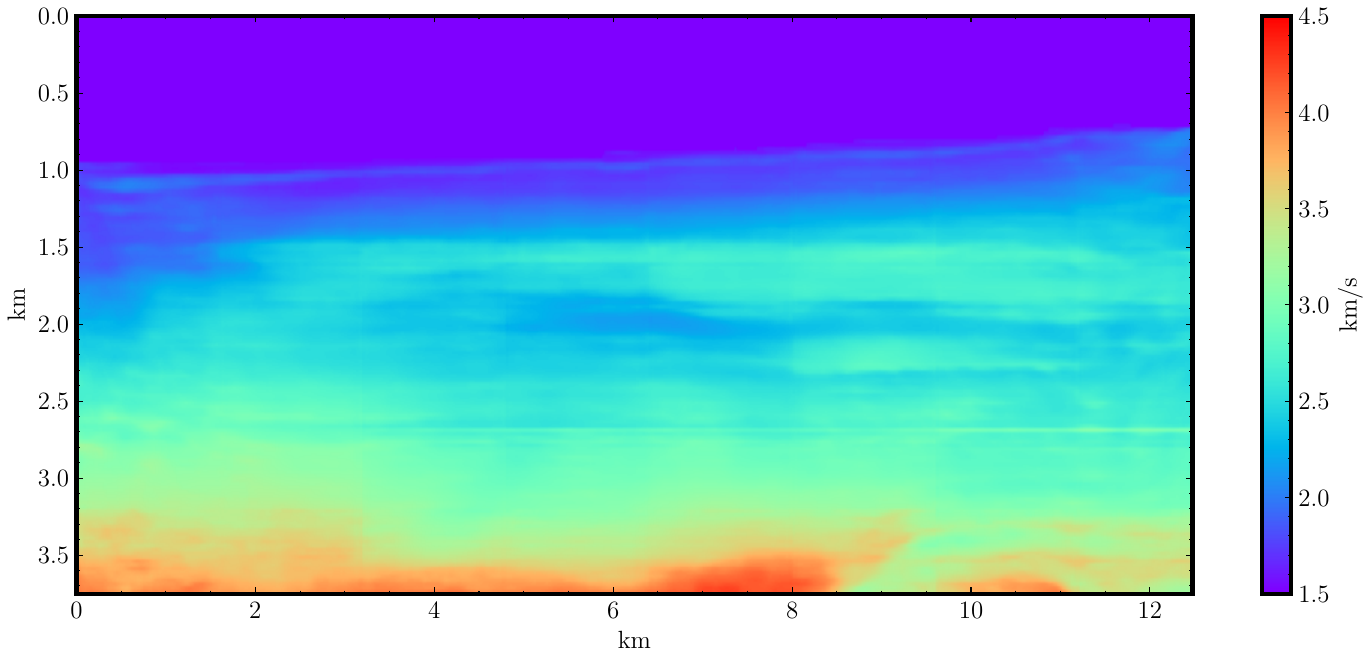}}

  \subfigure[Posterior mean \label{field_mean_seg}]{
    \includegraphics[width=0.4\columnwidth]{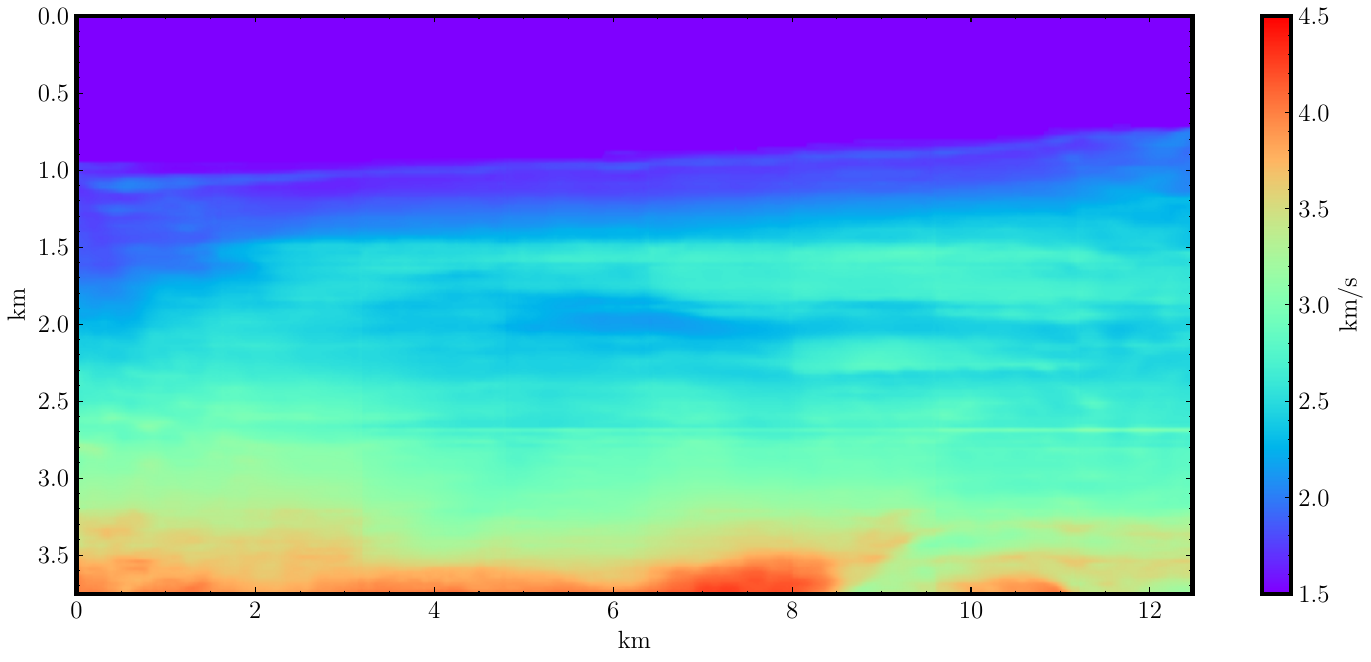}}
  \subfigure[Posterior standard deviation (diffusion)\label{field_std_seg}]{
    \includegraphics[width=0.4\columnwidth]{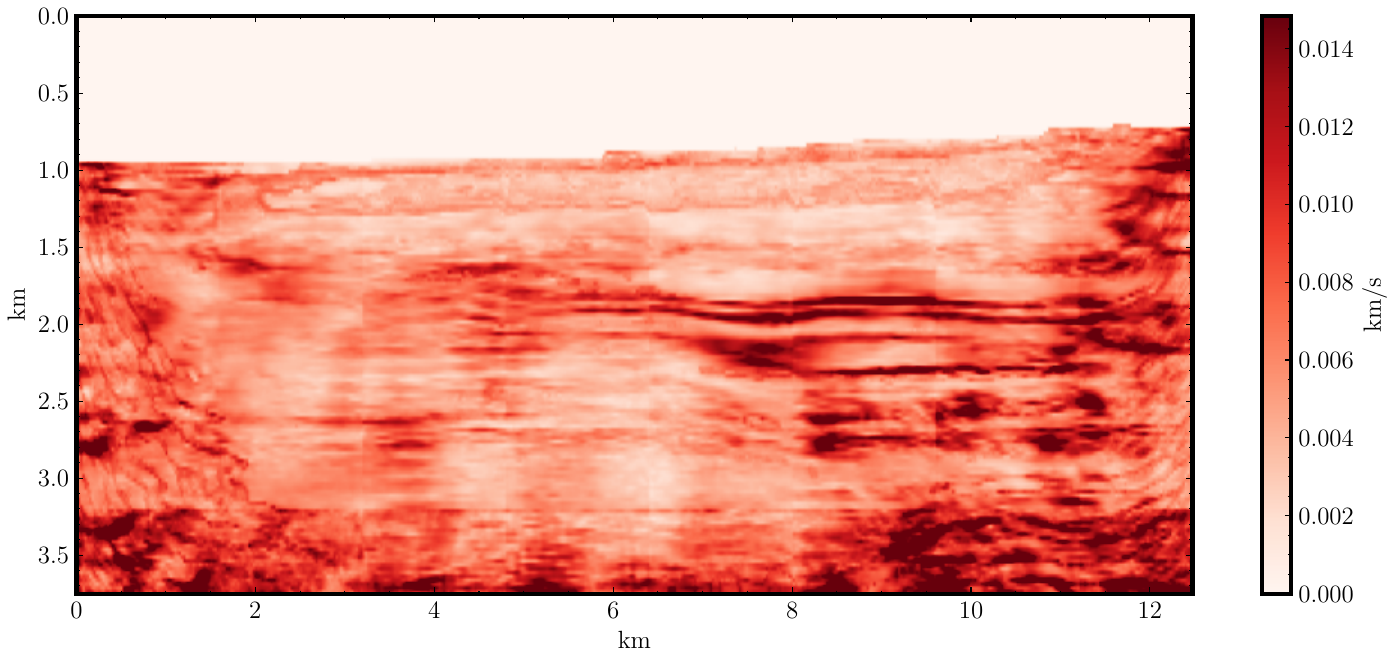}}

  \caption{2D towed-streamer field data with a diffusion prior trained on the \textsc{Realistic2D} velocity family and simultaneous-source likelihood guidance. (a) Initial model; (b) one posterior sample; (c) posterior mean; (d) posterior standard deviation.}
  \label{field_2d_seg}
\end{figure}

\begin{figure}
  \centering
  \subfigure[Initial velocity\label{field_init_svgd}]{
    \includegraphics[width=0.4\columnwidth]{Figures/field2d_init.pdf}}
  \subfigure[Posterior sample (SVGD)\label{field_sample_svgd}]{
    \includegraphics[width=0.4\columnwidth]{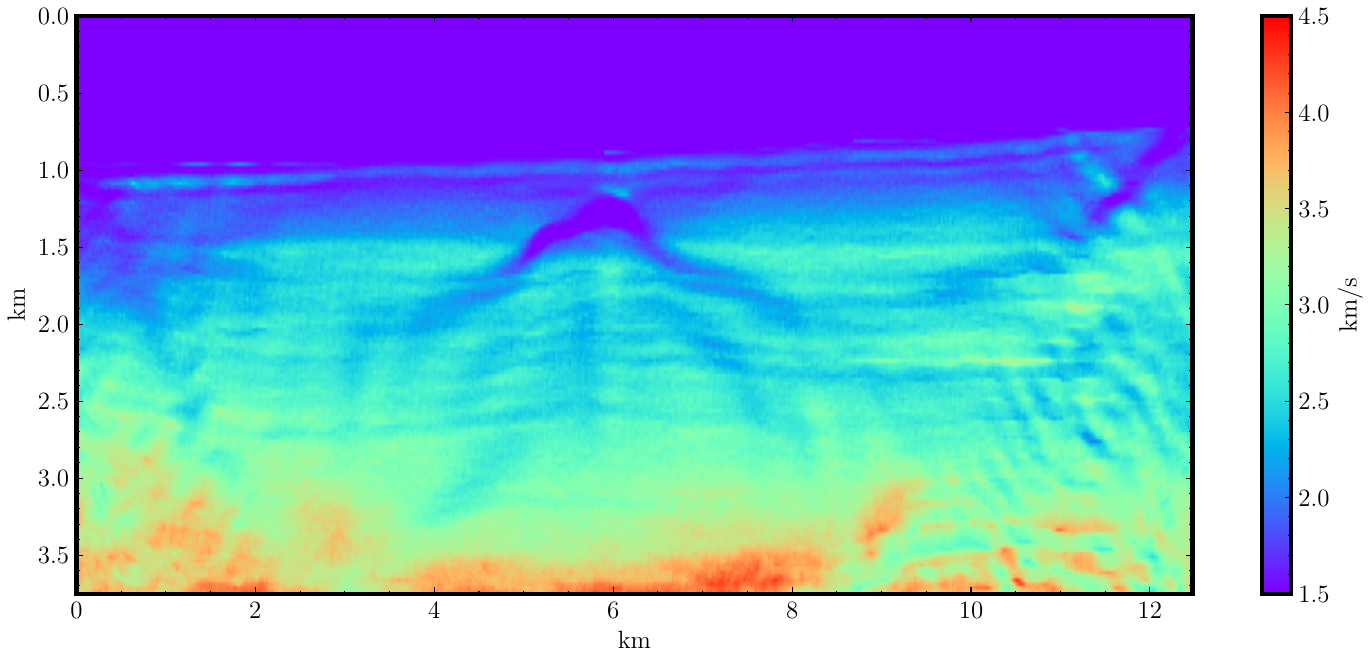}}

  \subfigure[Posterior mean (SVGD)\label{field_mean_svgd}]{
    \includegraphics[width=0.4\columnwidth]{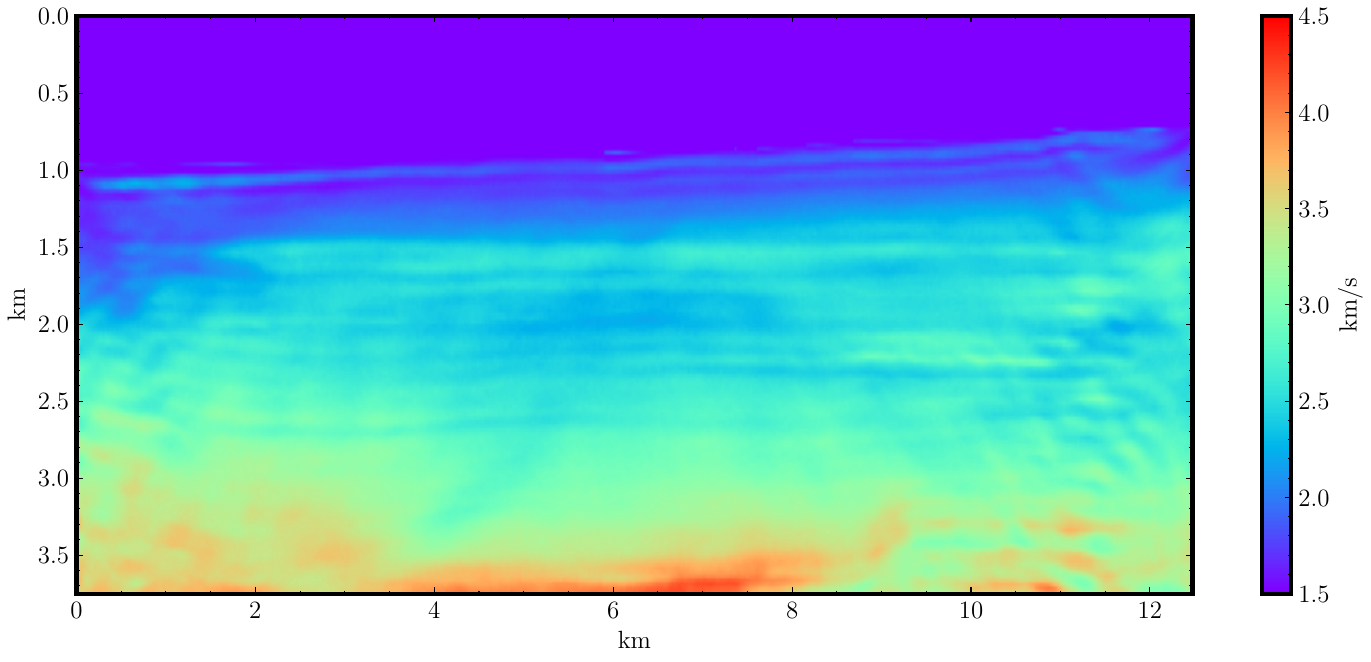}}
  \subfigure[Posterior standard deviation (SVGD)\label{field_std_svgd}]{
    \includegraphics[width=0.4\columnwidth]{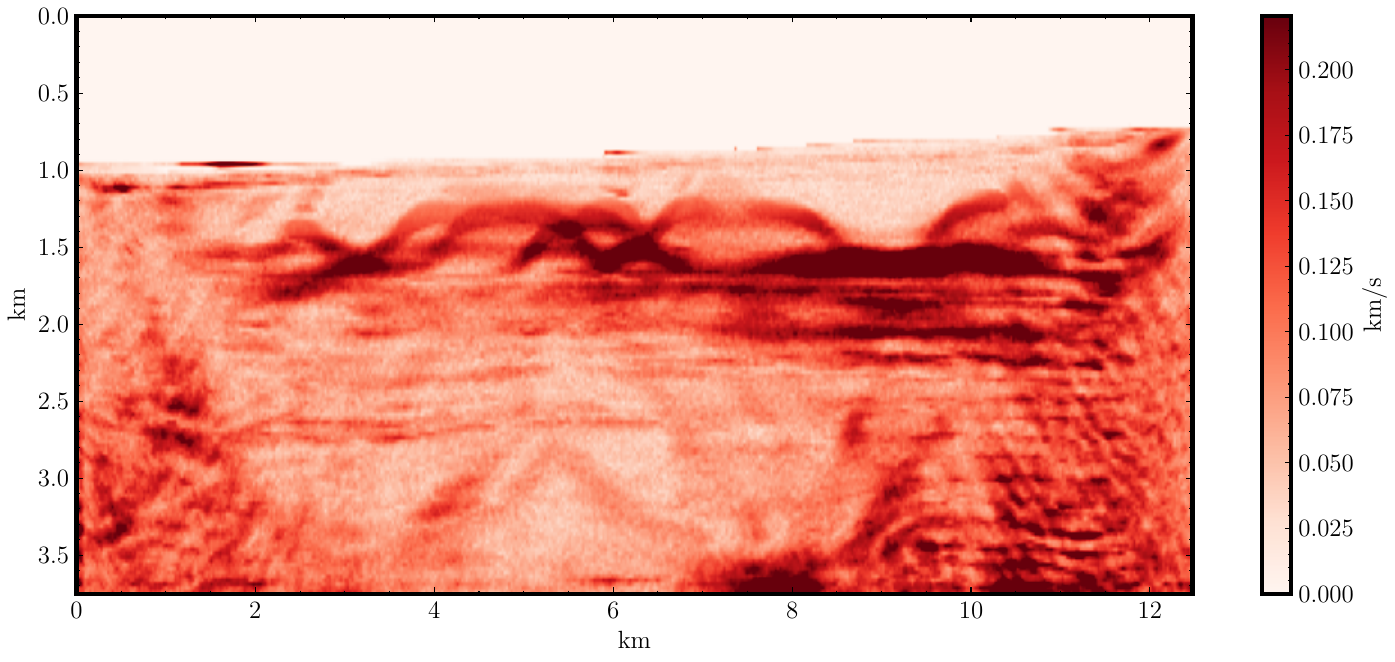}}

  \caption{2D towed-streamer field data with SVGD using conventional FWI gradients. (a) Initial model; (b) one posterior sample; (c) posterior mean; (d) posterior standard deviation.}
  \label{field_2d_svgd}
\end{figure}

We conduct two diffusion model posterior sampling utilizing two diffusion models trained on different training sets. Specifically, the two training sets provide a distinct vertical velocity resolution, with the \textsc{Random2D} possessing a much higher resolution than the \textsc{Realistic2D} training set. The vertical velocity resolution mismatch can also be observed in the generated diffusion posterior samples (Figures \ref{field_2d_random} and \ref{field_2d_seg}). More importantly, in line with our intuition, the two sets of posterior samples are in agreement in areas with good data illumination and differ in areas with poor data illumination. The SVGD posterior sample, on the other hand, shows a hint of a suboptimal convergence, judging by the anomalous low velocity layer in the shallow part of the model (Figure \ref{field_2d_svgd}b).

\begin{figure}
  \centering
  \subfigure[]{
    \includegraphics[width=\columnwidth]{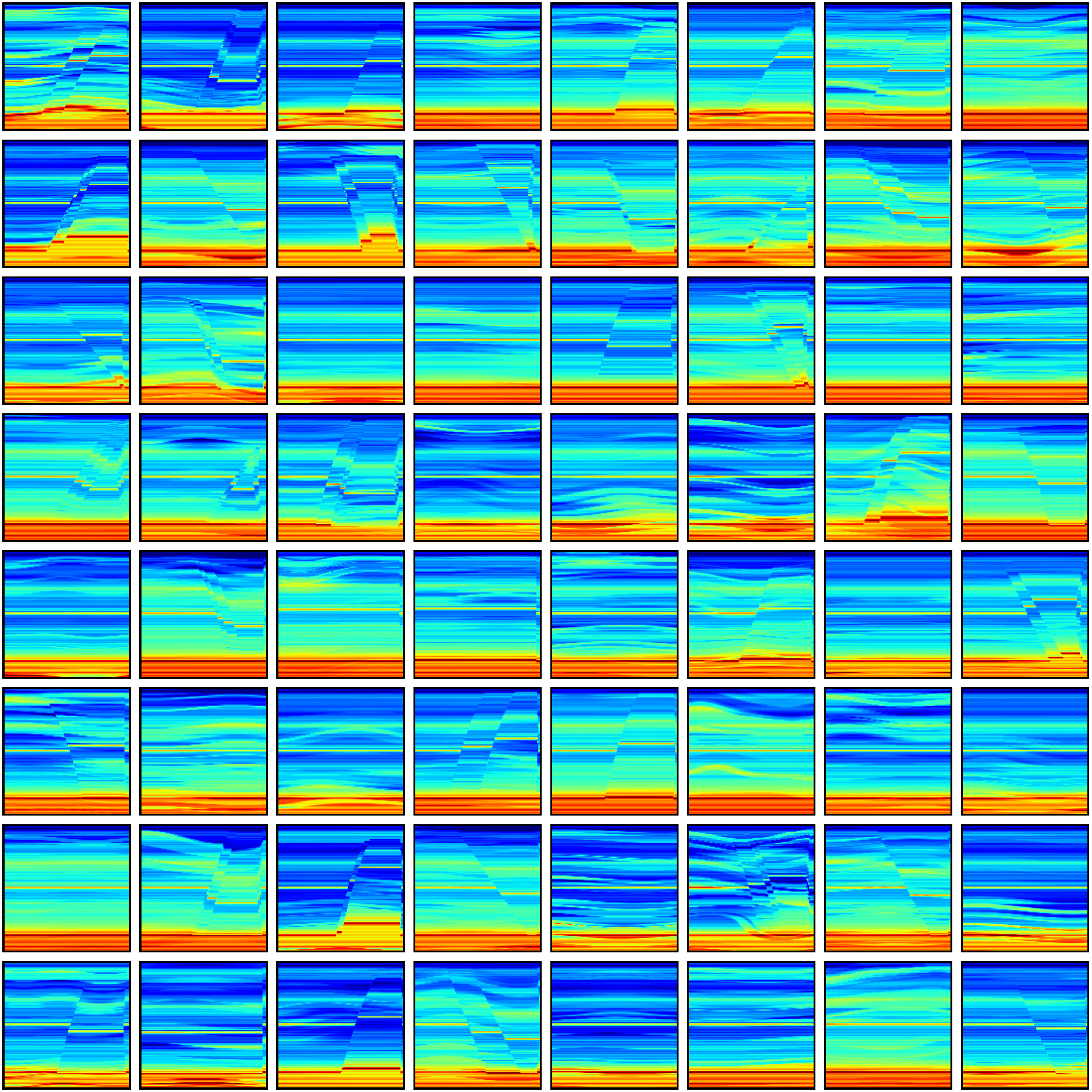}}
  \caption{Samples from the training (\textsc{Random2D}) dataset for the 2D diffusion model.}
  \label{data_random_2d}
\end{figure}

\subsection{3D ocean-bottom node synthetic data upscaling}

One of the main important features of our diffusion model inference is its ability to be data independent. Specifically, we argue that for large-scale high-dimensional problems, like FWI, deploying the diffusion model to solely work within the velocity model domain offers significant practicality. By solely working in the velocity model domain, the diffusion model is independent of any data-related requirements when dealing with different survey acquisitions. Furthermore, it also provides natural scalability to handle a larger velocity model size. In this example, we demonstrate that we can use a small 3D diffusion to handle a larger acquisition area.

\begin{figure}
  \centering
  \subfigure[Acquisition geometry (10\,km $\times$ 10\,km)\label{compass_10_mask}]{
    \includegraphics[width=0.4\columnwidth]{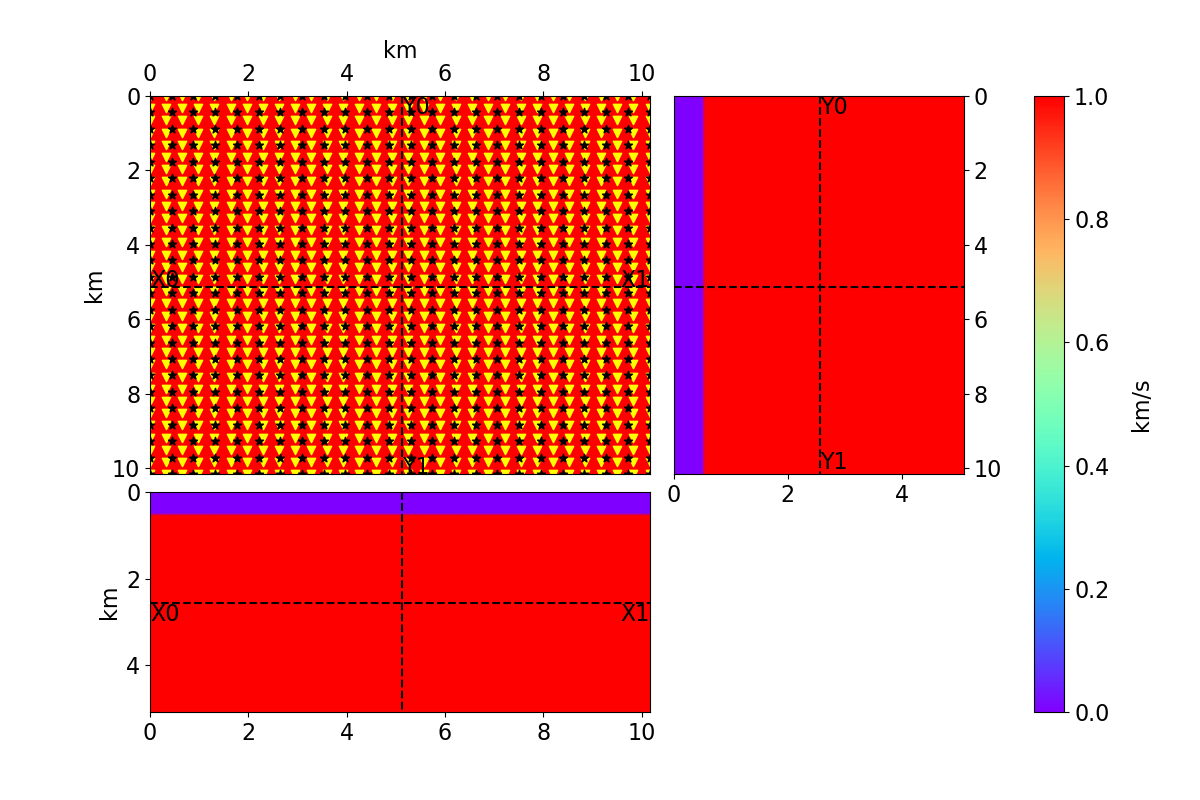}}
  \subfigure[Recorded data (three frequencies)\label{compass_10_dobs}]{
    \includegraphics[width=0.4\columnwidth]{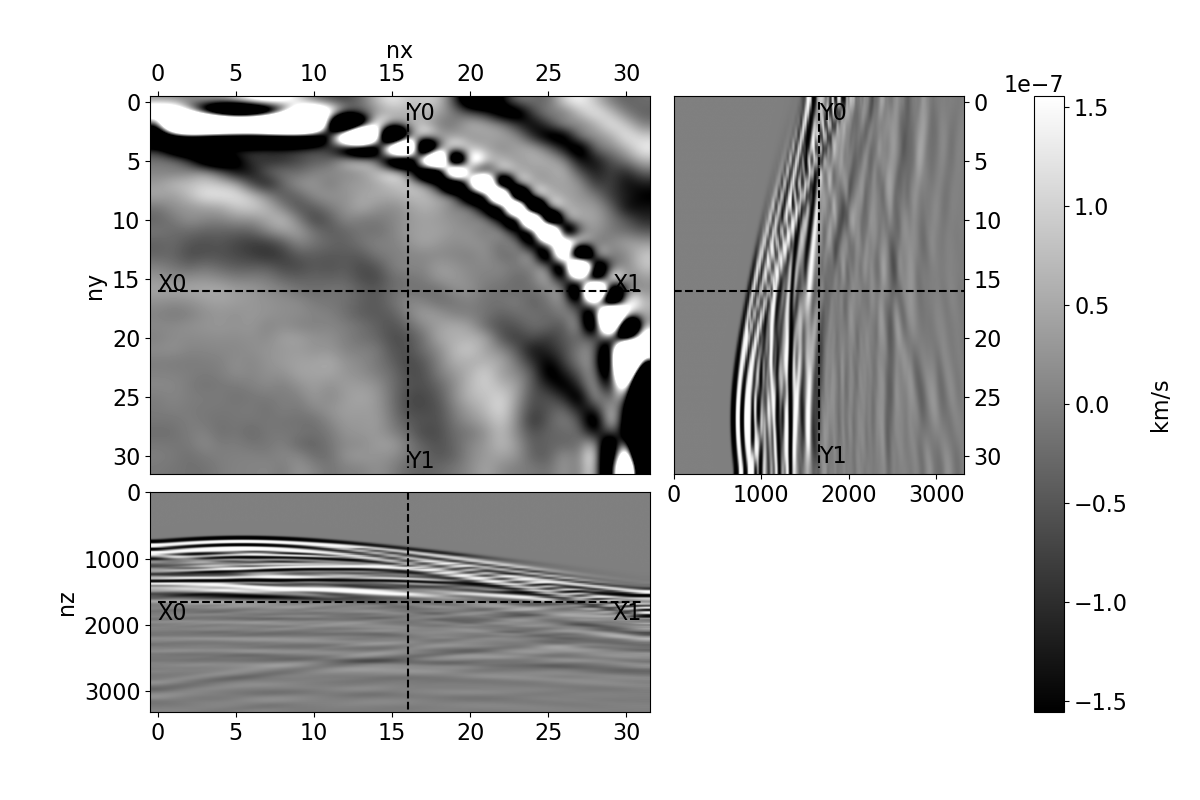}}

  \subfigure[Initial velocity\label{compass_10_init}]{
    \includegraphics[width=0.4\columnwidth]{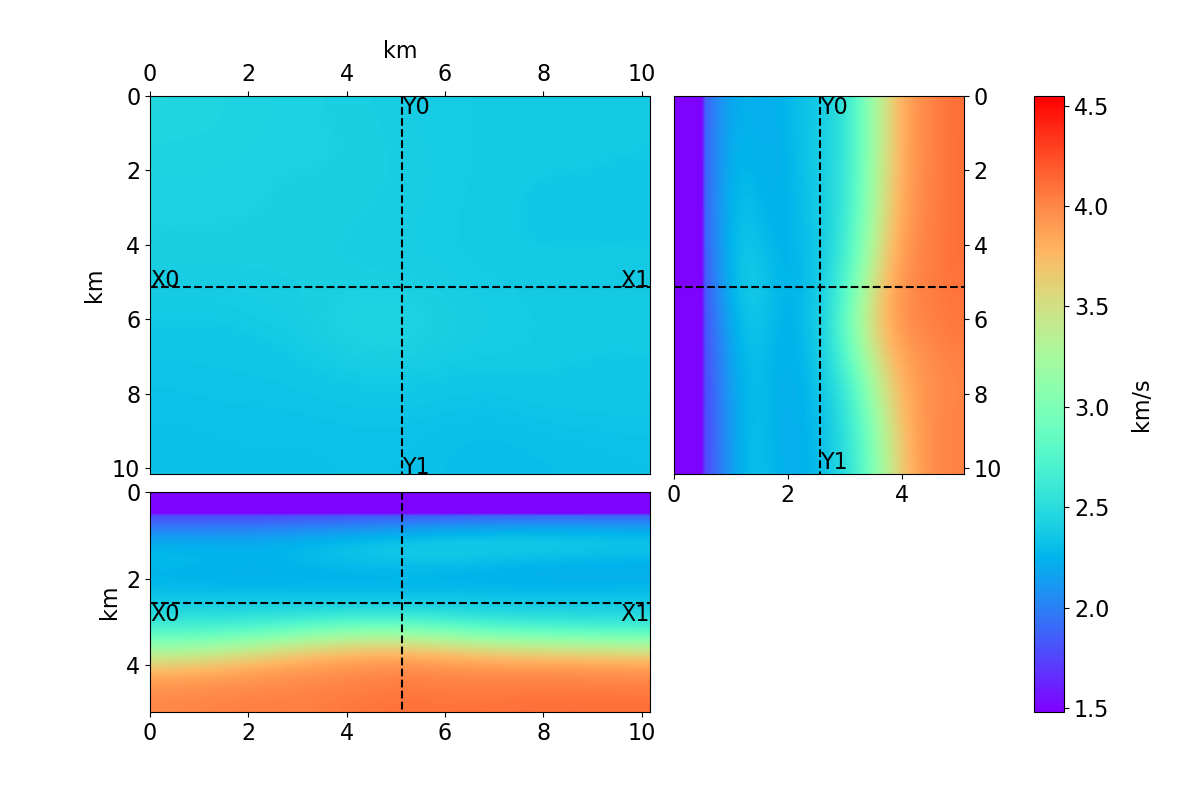}}
  \subfigure[True velocity\label{compass_10_true}]{
    \includegraphics[width=0.4\columnwidth]{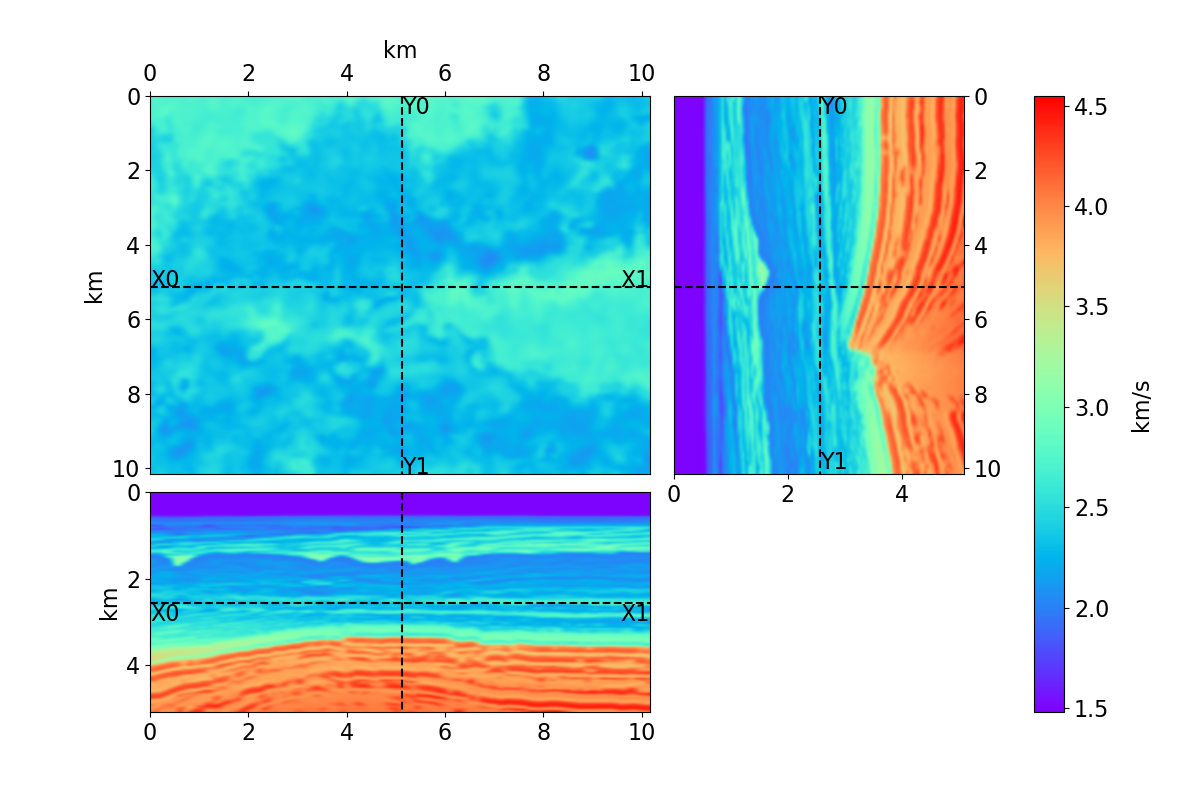}}

  \subfigure[Multi-source FWI (without diffusion guidance)\label{compass_10_conv}]{
    \includegraphics[width=0.4\columnwidth]{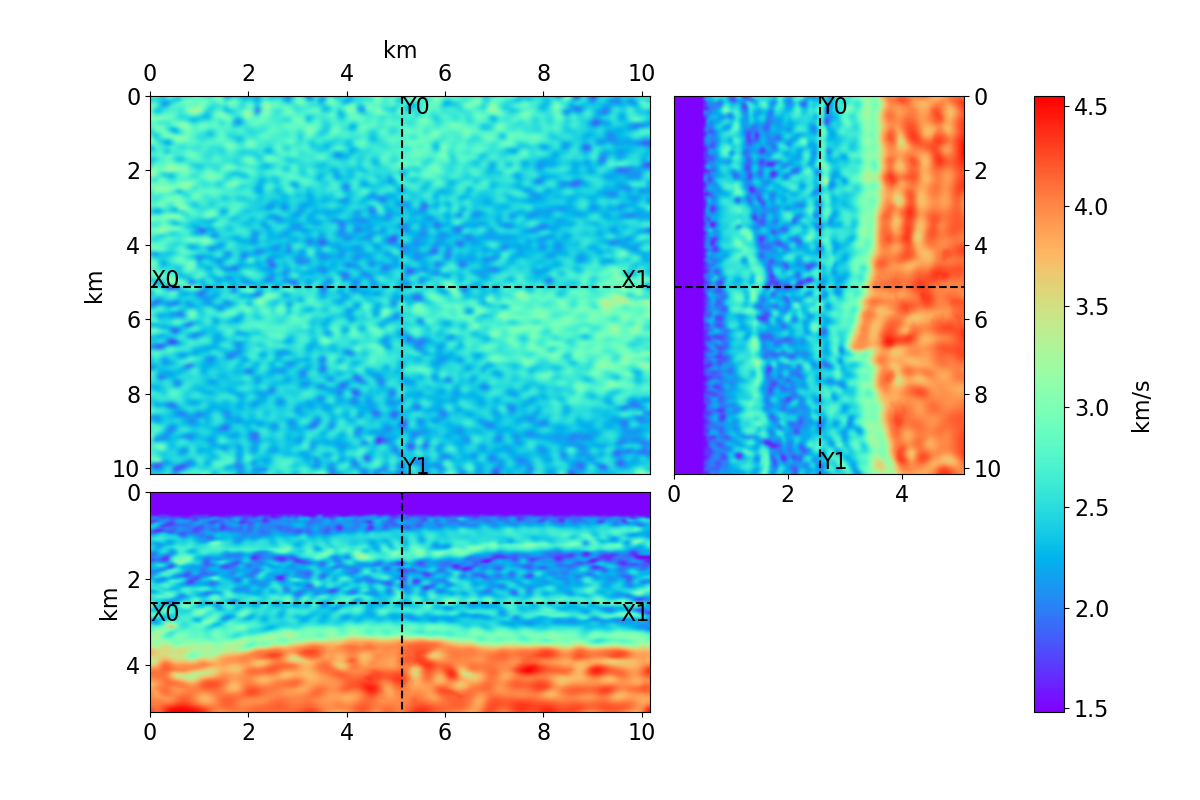}}
  \subfigure[Multi-source FWI (with diffusion guidance)\label{compass_10_diff}]{
    \includegraphics[width=0.4\columnwidth]{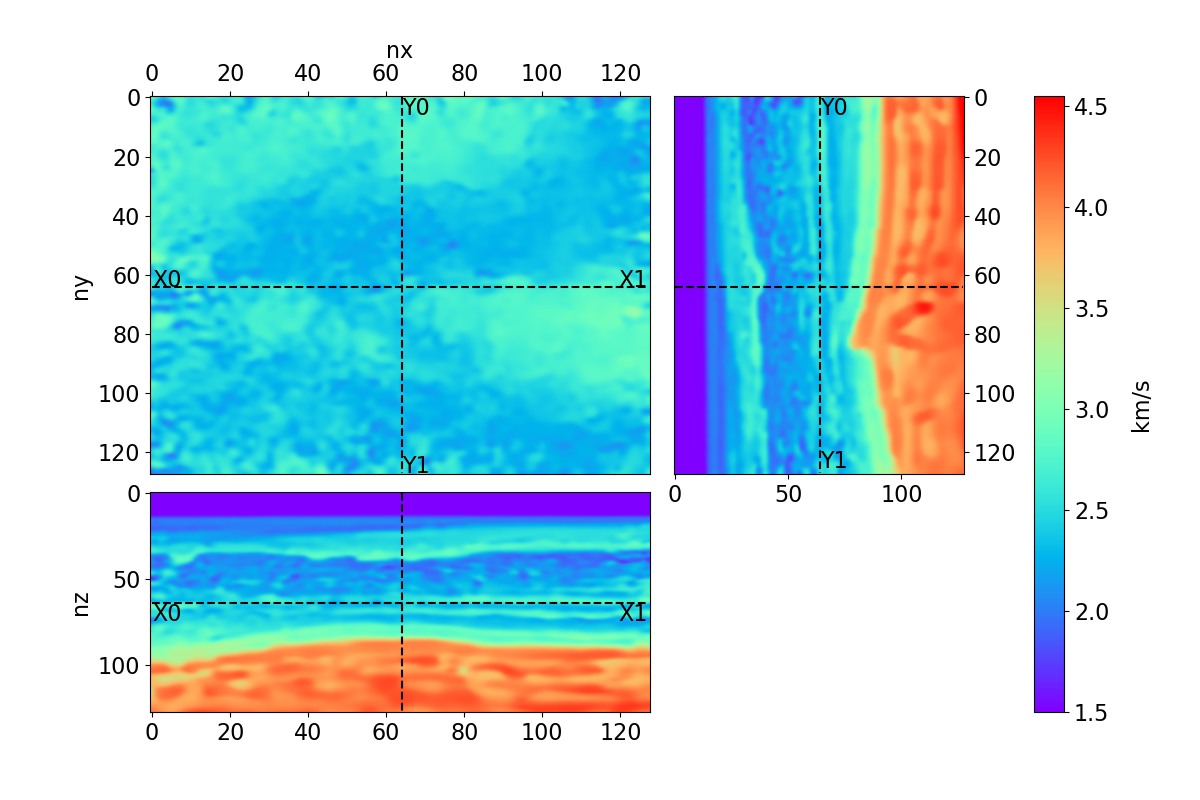}}

  \caption{BG Compass synthetic (3D OBN), 10\,km $\times$ 10\,km area. (a) Acquisition geometry; (b) recorded data (three frequencies); (c) initial model; (d) true model; (e) multi-source FWI without diffusion guidance; (f) multi-source FWI with diffusion guidance.}
  \label{compass_10}
\end{figure}

\begin{figure}
  \centering
  \subfigure[Acquisition geometry (20\,km $\times$ 20\,km)\label{compass_20_mask}]{
    \includegraphics[width=0.4\columnwidth]{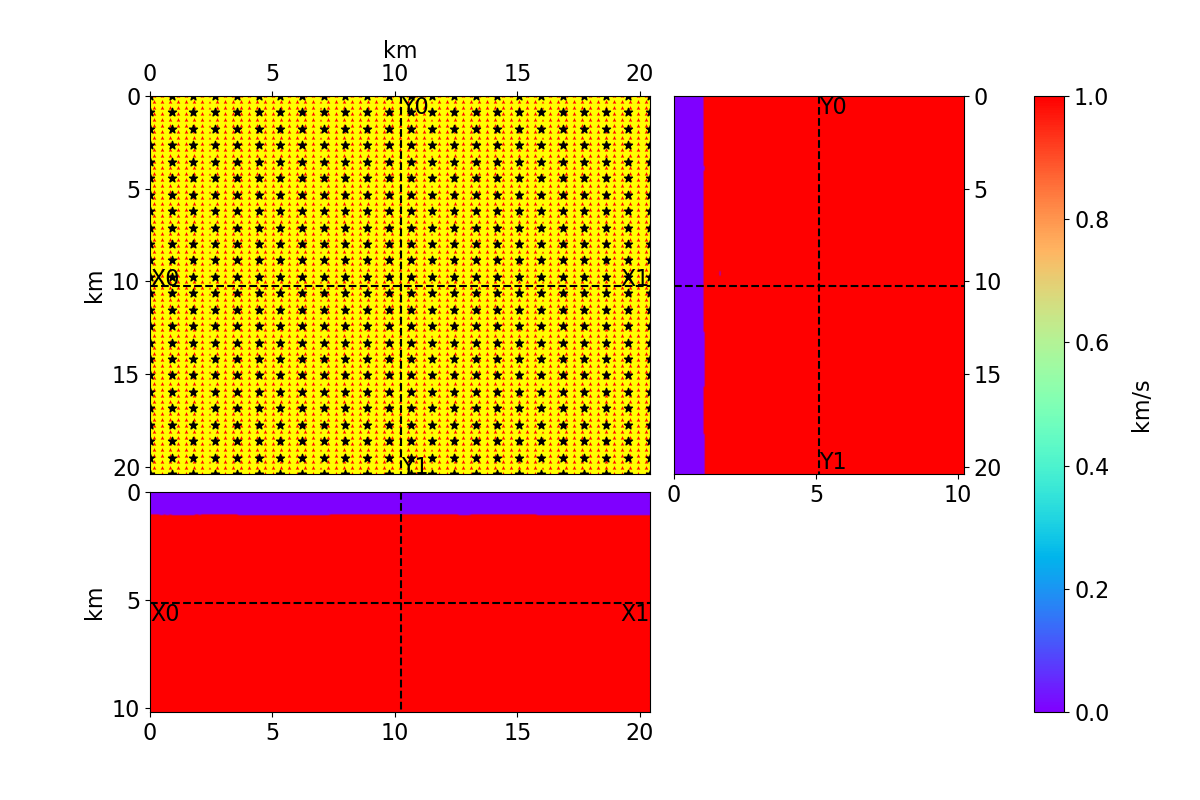}}
  \subfigure[Recorded data (three frequencies)\label{compass_20_dobs}]{
    \includegraphics[width=0.4\columnwidth]{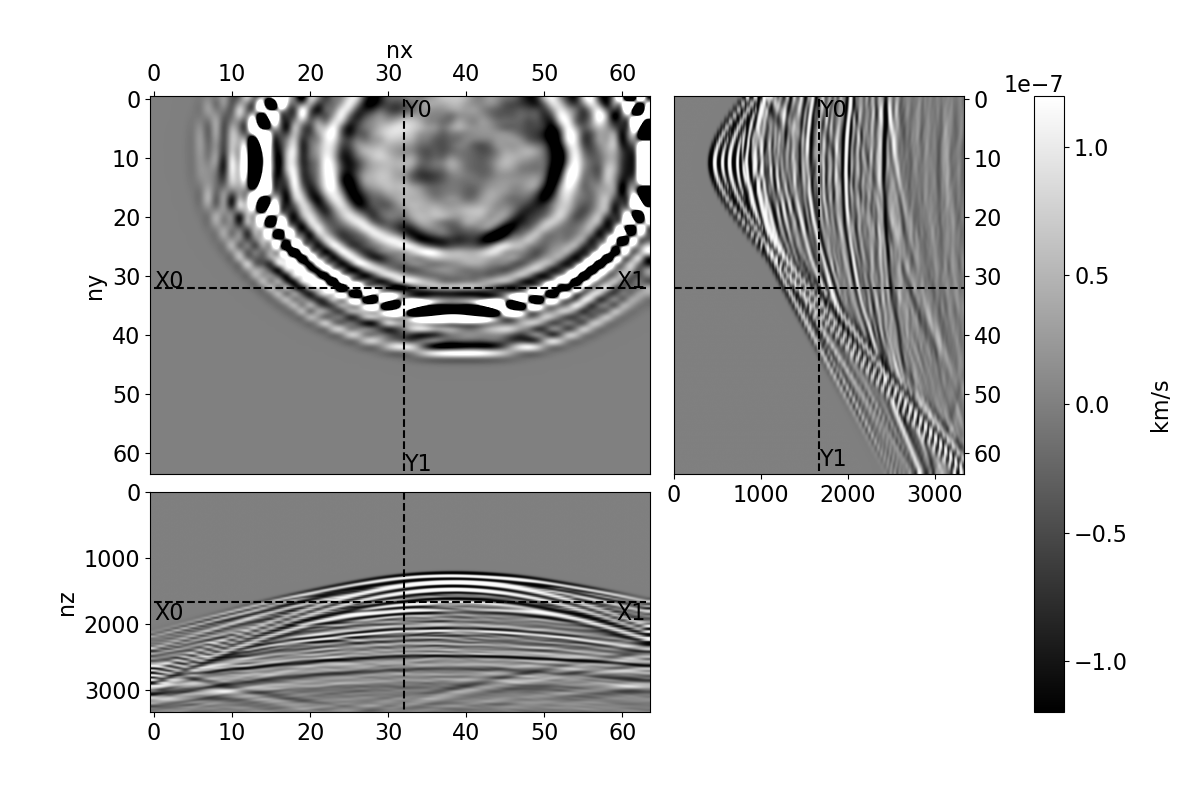}}

  \subfigure[Initial velocity\label{compass_20_init}]{
    \includegraphics[width=0.4\columnwidth]{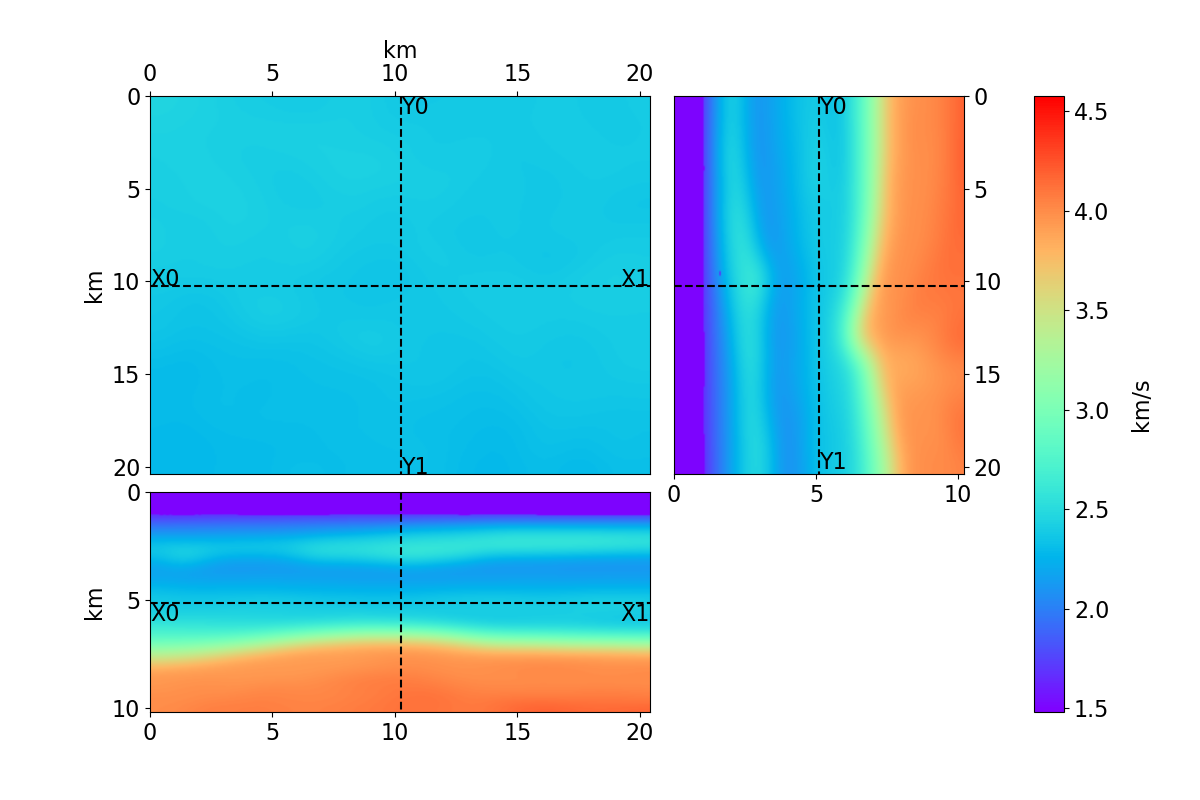}}
  \subfigure[True velocity\label{compass_20_true}]{
    \includegraphics[width=0.4\columnwidth]{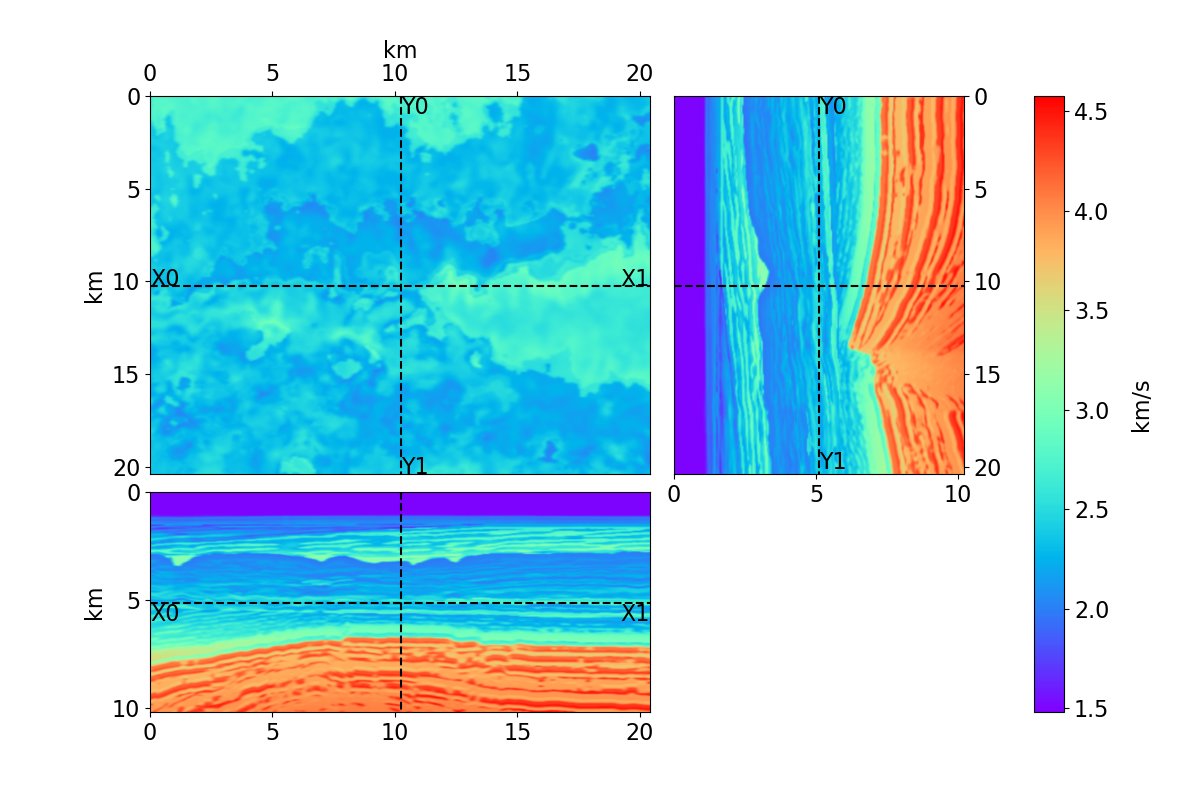}}

  \subfigure[Multi-source FWI (without diffusion guidance)\label{compass_20_conv}]{
    \includegraphics[width=0.4\columnwidth]{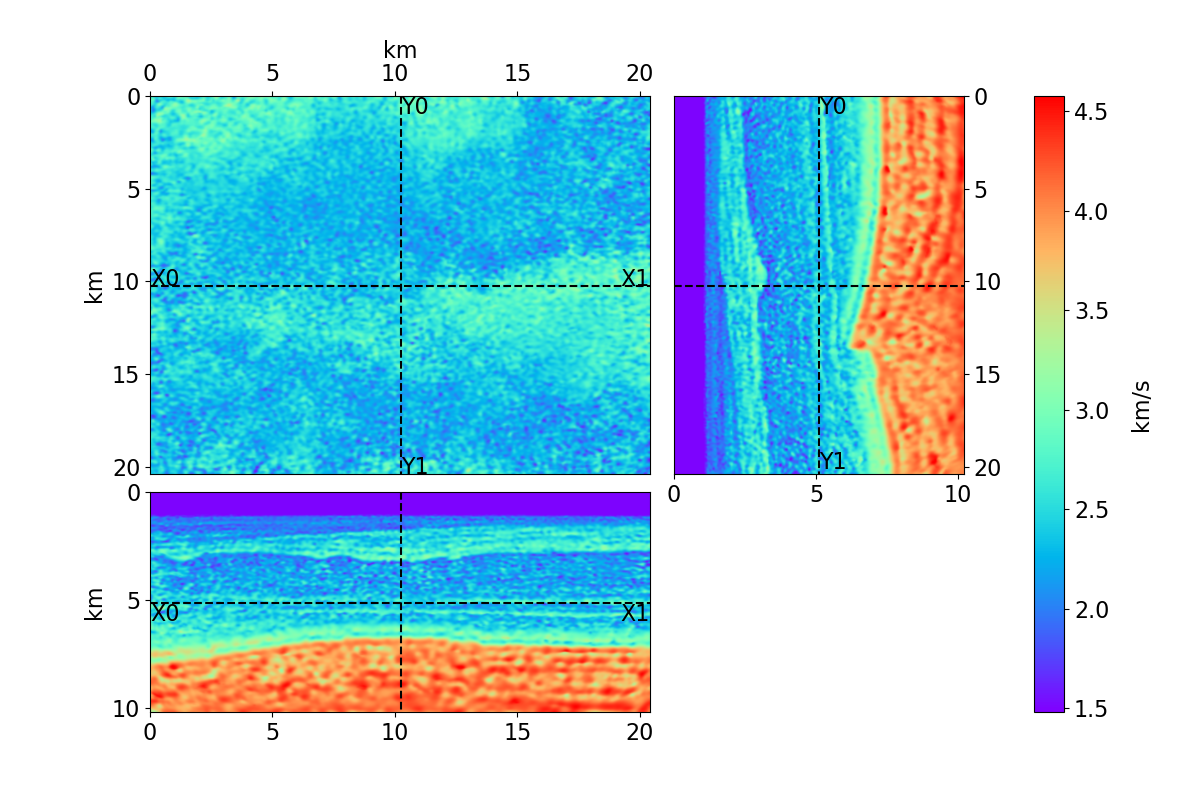}}
  \subfigure[Multi-source FWI (with diffusion guidance)\label{compass_20_diff}]{
    \includegraphics[width=0.4\columnwidth]{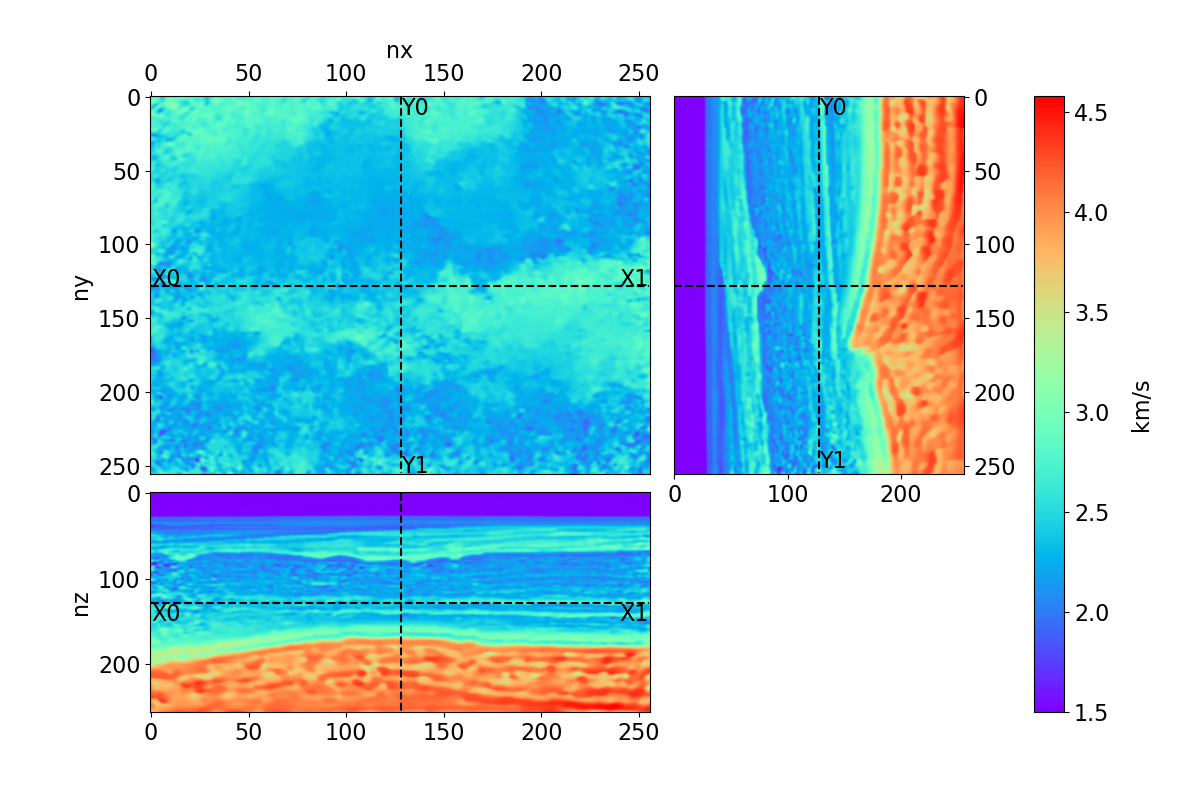}}

  \caption{BG Compass synthetic (3D OBN), 20\,km $\times$ 20\,km area. (a) Acquisition geometry; (b) recorded data (three frequencies); (c) initial model; (d) true model; (e) multi-source FWI without diffusion guidance; (f) multi-source FWI with diffusion guidance.}
  \label{compass_20}
\end{figure}

\begin{figure}
  \centering
  \subfigure[Acquisition geometry (30\,km $\times$ 30\,km)\label{compass_30_mask}]{
    \includegraphics[width=0.4\columnwidth]{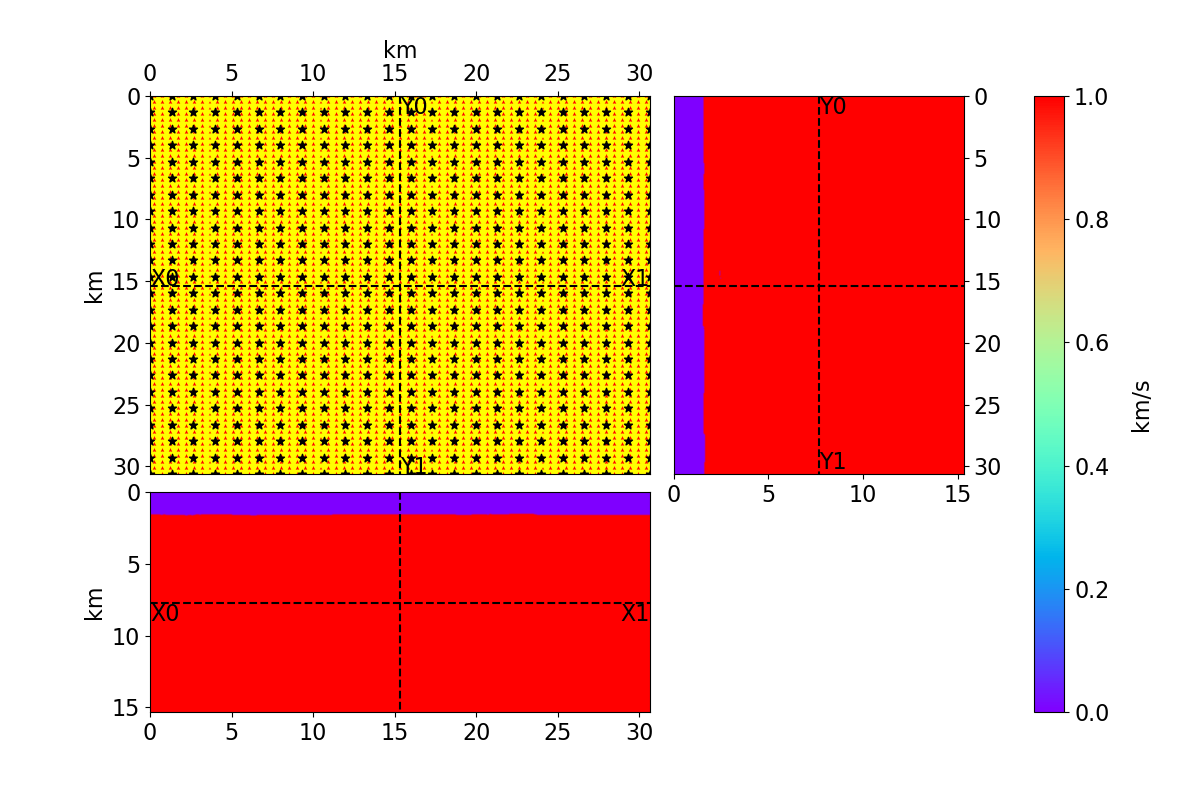}}
  \subfigure[Recorded data (three frequencies)\label{compass_30_dobs}]{
    \includegraphics[width=0.4\columnwidth]{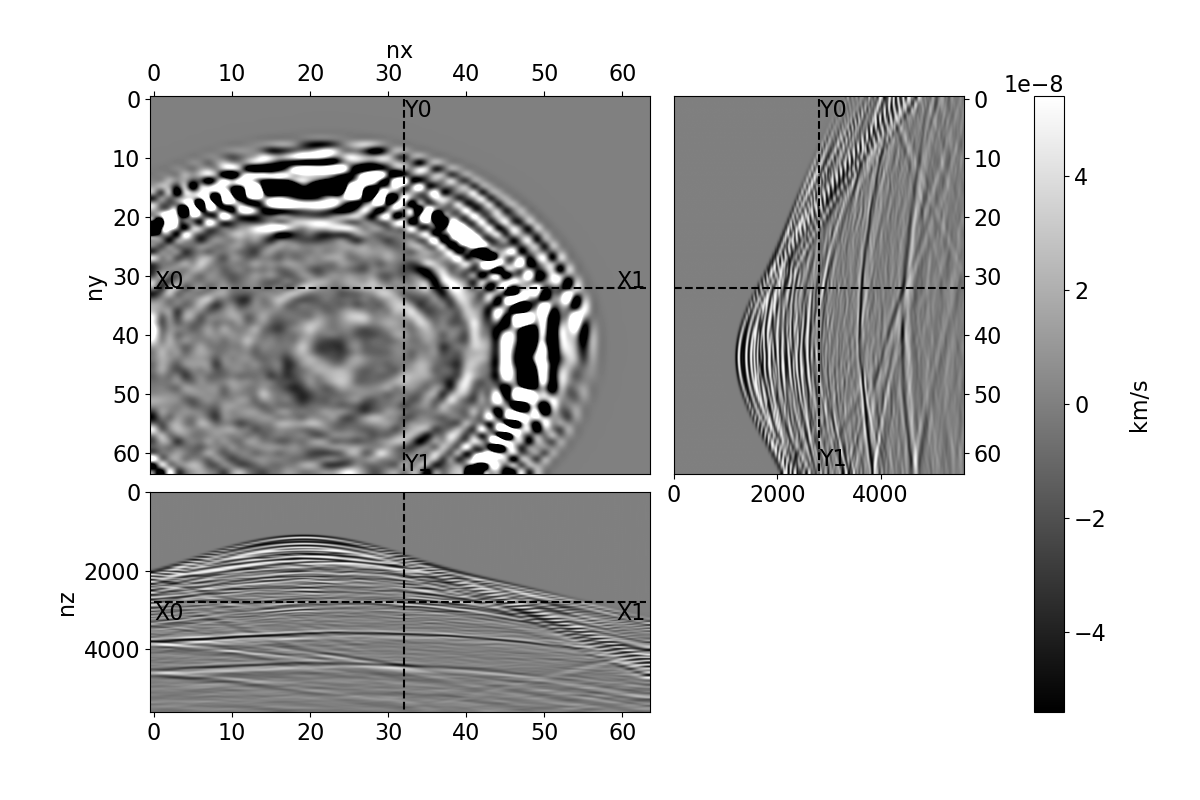}}

  \subfigure[Initial velocity\label{compass_30_init}]{
    \includegraphics[width=0.4\columnwidth]{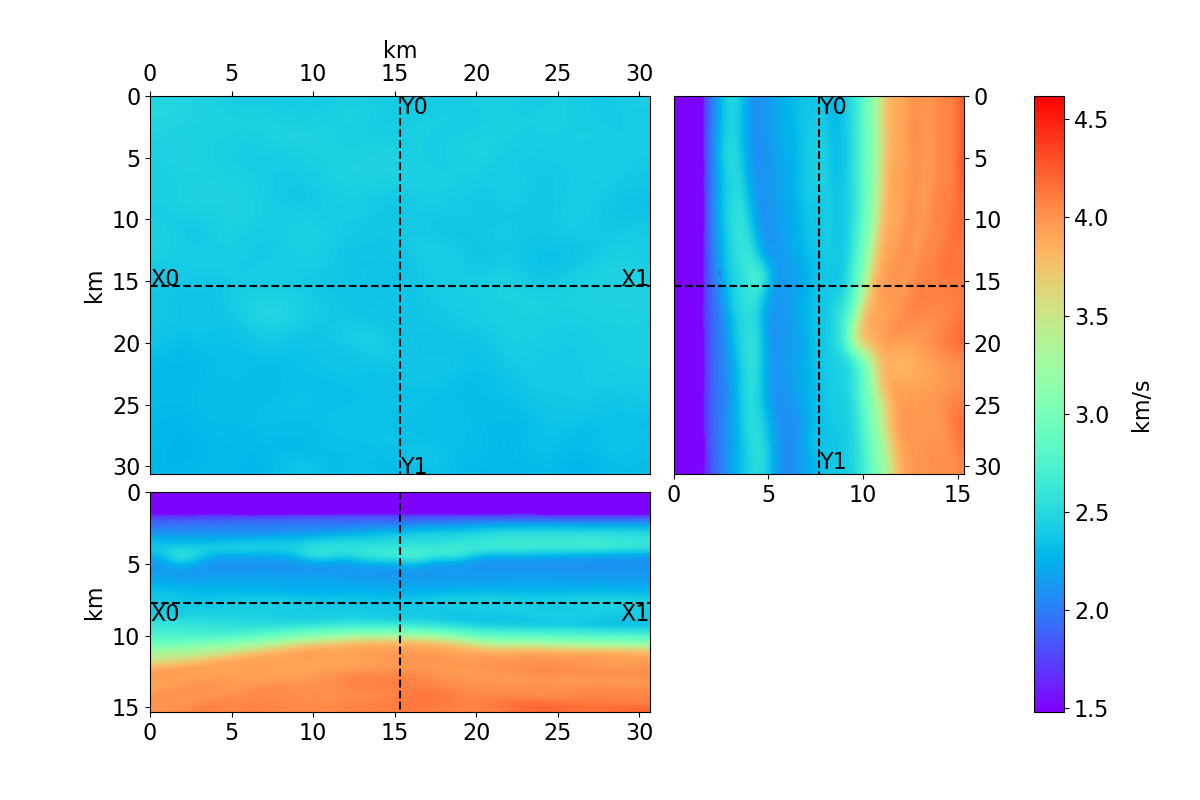}}
  \subfigure[True velocity\label{compass_30_true}]{
    \includegraphics[width=0.4\columnwidth]{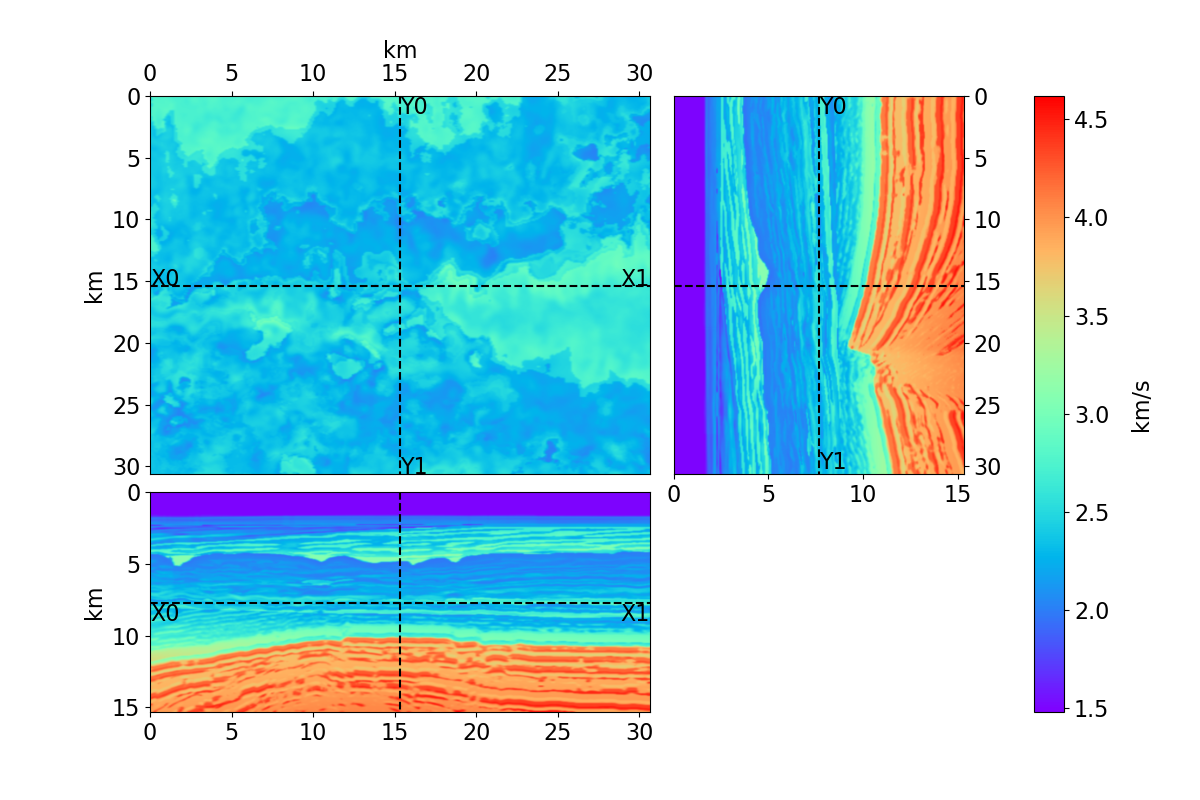}}

  \subfigure[Multi-source FWI (without diffusion guidance)\label{compass_30_conv}]{
    \includegraphics[width=0.4\columnwidth]{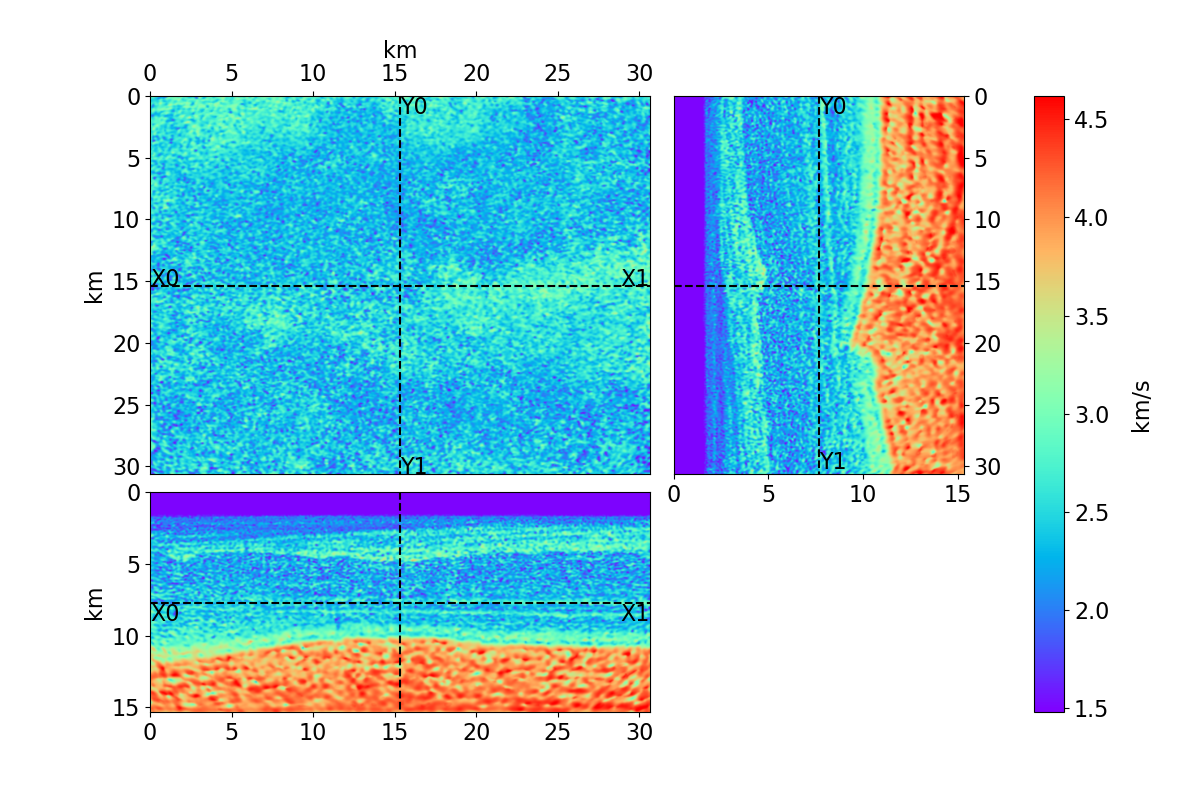}}
  \subfigure[Multi-source FWI (with diffusion guidance)\label{compass_30_diff}]{
    \includegraphics[width=0.4\columnwidth]{Figures/synthetic3d_20_Velocity_0_0.png}}

  \caption{BG Compass synthetic (3D OBN), 30\,km $\times$ 30\,km area. (a) Acquisition geometry; (b) recorded data (three frequencies); (c) initial model; (d) true model; (e) multi-source FWI without diffusion guidance; (f) multi-source FWI with diffusion guidance.}
  \label{compass_30}
\end{figure}

To do so, we setup three different acquisition areas with details described in Table \ref{tab:acq-areas-merged}. We devise the same patching strategy as in the 2D case to mitigate the mismatch between the diffusion input and the velocity model size. We use a stride of size 32 on both of the 10$\times$10 and 20$\times$20 km sq experiments, while we use non-overlapping patching for the 30x30 km sq experiment. Shown in Figures \ref{compass_10}, \ref{compass_20}, and \ref{compass_30}, the diffusion model not only manages to scale for larger velocity size, but it also removes some of the source-related artefacts when doing simultaneous-source FWI experiments.


\begin{table*}
  \centering
  \caption{Acquisition parameters for the three 3D OBN areas.}
  \label{tab:acq-areas-merged}
  \small
  \setlength{\tabcolsep}{8pt}
  \renewcommand{\arraystretch}{1.15}
  \begin{tabular}{lccc}
    \toprule
    \textbf{Parameter} & \textbf{Area A} & \textbf{Area B} & \textbf{Area C} \\
    \midrule
    Survey footprint (km) & $10 \times 10$ & $20 \times 20$ & $30 \times 30$ \\
    Model grid $(n_x,n_y,n_z)$ & $128,\,128,\,128$ & $256,\,256,\,256$ & $384,\,384,\,384$ \\
    Grid spacing $(\Delta x,\Delta y,\Delta z)$ & $80\,\mathrm{m},\,80\,\mathrm{m},\,40\,\mathrm{m}$ & $80\,\mathrm{m},\,80\,\mathrm{m},\,40\,\mathrm{m}$ & $80\,\mathrm{m},\,80\,\mathrm{m},\,40\,\mathrm{m}$ \\
    Depth extent $n_z\Delta z$ & $5.12\,\mathrm{km}$ & $5.12\,\mathrm{km}$ & $15.36\,\mathrm{km}$ \\
    Sources $(n_s)$ & $576$ & $576$ & $576$ \\
    Source spacing $d_s$ & $0.42\,\mathrm{km}$ & $0.42\,\mathrm{km}$ & $1.28\,\mathrm{km}$ \\
    Receivers $(n_r)$ & $1024$ & $4096$ & $4096$ \\
    Receiver spacing $d_r$ & $320\,\mathrm{m}$ & $320\,\mathrm{m}$ & $480\,\mathrm{m}$ \\
    \textbf{Number of supergathers $(m)$} & \textbf{4} & \textbf{4} & \textbf{9} \\
    Samples per trace $(n_t)$ & $4000$ & $4000$ & $5000$ \\
    Sampling interval $(\Delta t)$ & $0.003\,\mathrm{s}$ & $0.003\,\mathrm{s}$ & $0.004\,\mathrm{s}$ \\
    Record length $n_t\Delta t$ & $12\,\mathrm{s}$ & $12\,\mathrm{s}$ & $20\,\mathrm{s}$ \\
    Reference frequency & $3\,\mathrm{Hz}$ & $3\,\mathrm{Hz}$ & $3\,\mathrm{Hz}$ \\
    \bottomrule
  \end{tabular}
\end{table*}

\subsection{Comparison with a variational inference baseline}

We benchmark our diffusion–likelihood sampler against a strong particle-based variational baseline (SVGD without a learned prior and using standard, non-encoded shots) to understand where the gains come from: data fit versus model accuracy, computational cost, and posterior calibration. The key design choices in our method—using an unconditional diffusion prior to regularize plausible geology, coupling it with simultaneous-source (encoded-shot) likelihood gradients, and inserting a light stochastic refinement between diffusion levels—aim to reduce forward/adjoint counts while avoiding the mode-seeking behavior often observed in practical VI on FWI. This comparison, therefore, probes whether a generative prior plus encoded data-fidelity actually translates to better inversions at the field scale, not just cleaner samples. 

On the 2D synthetics (Overthrust, Salt, Arid), our sampler consistently lowers both model-space error and data-space misfit relative to SVGD. For Overthrust, we observe a large reduction in velocity RMSE ($\approx$54\%) and an order-of-magnitude drop in NRMS ($\approx$87\%), indicating that improvements are not confined to visual plausibility but carry through to data-fitting agreement. The salt model shows smaller but still material gains (RMSE $\approx$23\%, NRMS $\approx$20\%), while Arid exhibits modest model-space improvement yet a clear data-space advantage, which we attribute to better handling of illumination gaps. In field data, the improvements are more conservative—as expected—but remain consistent across correlation, time-shift, envelope, and band-limited spectral misfits, suggesting the method transfers beyond controlled synthetics. Taken together, these patterns imply that the diffusion prior primarily curbs implausible updates (stabilizing the model error), and the encoded-shot likelihood sharpens data conformity without incurring the full multi-source cost. 

From a cost perspective, simultaneous sources provide unbiased likelihood gradients whose variance is well behaved, so that one PDE solve effectively stands in for many; in practice, we reduce the number of forward/adjoint solves approximately in proportion to the supergather size. Because the sampler also operates level-wise—predict, refine briefly under the physics, re-noise to decorrelate, then proceed—we obtain parallelism over both samples and source batches. In a head-to-head comparison with SVGD (no prior, non-encoded shots), this translates into lower wall-clock time for comparable or better misfit, and a strictly better accuracy–cost trade-off in the regimes we tested. 

Uncertainty calibration is where the stochastic refinement matters the most. Ablation tests show that purely deterministic guidance (no stochastic refinement and no level re-noising) tends to underestimate posterior variance even when mean models are visually plausible. Introducing short, noise-aware refinements at each diffusion level and re-noising to decouple levels restores dispersion that tracks the true data information content (illumination and noise), yielding posterior means/variances that pass standard posterior-predictive checks more reliably than the SVGD baseline. 

Finally, we note limits and failure modes. The diffusion prior can, in principle, bias solutions toward the training distribution, which acts as prior; however, in our tests, the encoded-shot likelihood consistently counteracts this by pulling samples toward data-consistent modes, and the re-noising prevents premature collapse. Conversely, SVGD’s behavior remains sensitive to kernel bandwidth, particle count, and step-size schedules; without careful encoded-shot scaling, it inherits full-shot costs while still tending to be underestimated. Overall, the evidence indicates that combining a learned prior with encoded data-fidelity and light stochastic refinement yields better data fit, lower model error, improved calibration, and materially lower computational burden than a variational baseline tuned conventionally.

\begin{landscape}
\begin{table*}
\centering
\caption{Velocity–model metrics (mean$\pm$std) for synthetic datasets. Bold indicates best (lower is better except for Pearson $r$).}
\label{tab:model-metrics}
\begin{tabular}{lcccccc}
\toprule
\textbf{Metric}
& \makecell{\textbf{Overthrust}\\SVGD}
& \makecell{\textbf{Overthrust}\\Ours}
& \makecell{\textbf{Salt}\\SVGD}
& \makecell{\textbf{Salt}\\Ours}
& \makecell{\textbf{Arid}\\SVGD}
& \makecell{\textbf{Arid}\\Ours}
\\ \midrule
RMSE (m/s) $\downarrow$
& $377.550 \pm 4.338$ & \best{$172.312 \pm 2.436$}
& $226.391 \pm 4.998$ & \best{$175.151 \pm 2.178$}
& $363.340 \pm 1.059$ & \best{$346.395 \pm 5.464$}
\\
NRMSE ($v_{\max}$) $\downarrow$
& $0.063 \pm 0.001$ & \best{$0.029 \pm 0.000$}
& $0.051 \pm 0.001$ & \best{$0.039 \pm 0.000$}
& $0.058 \pm 0.000$ & \best{$0.055 \pm 0.001$}
\\
MAE (m/s) $\downarrow$
& $305.181 \pm 3.616$ & \best{$114.143 \pm 1.582$}
& $137.605 \pm 3.152$ & \best{$87.586 \pm 2.519$}
& $236.561 \pm 1.262$ & \best{$208.809 \pm 3.322$}
\\
relL2 $\downarrow$
& $0.082 \pm 0.001$ & \best{$0.038 \pm 0.001$}
& $0.092 \pm 0.002$ & \best{$0.071 \pm 0.001$}
& $0.086 \pm 0.000$ & \best{$0.082 \pm 0.001$}
\\
Pearson $r$ $\uparrow$
& $0.941 \pm 0.002$ & \best{$0.989 \pm 0.000$}
& $0.943 \pm 0.003$ & \best{$0.965 \pm 0.001$}
& $0.954 \pm 0.000$ & \best{$0.960 \pm 0.001$}
\\
Grad-MAE $\downarrow$
& $126.777 \pm 1.702$ & \best{$59.485 \pm 0.677$}
& $83.605 \pm 1.070$  & \best{$51.940 \pm 0.525$}
& $110.318 \pm 0.874$ & \best{$106.296 \pm 0.657$}
\\
Spec-relL2 $\downarrow$
& $0.051 \pm 0.000$ & \best{$0.024 \pm 0.000$}
& $0.055 \pm 0.002$ & \best{$0.038 \pm 0.001$}
& $0.055 \pm 0.000$ & \best{$0.047 \pm 0.001$}
\\
\bottomrule
\end{tabular}
\end{table*}

\begin{table*}
\centering
\caption{Data–domain metrics (mean$\pm$std) for synthetic datasets. Bold is best (lower is better except Trace corr).}
\label{tab:data-metrics}
\begin{tabular}{lcccccc}
\toprule
\textbf{Metric}
& \makecell{\textbf{Overthrust}\\SVGD}
& \makecell{\textbf{Overthrust}\\Ours}
& \makecell{\textbf{Salt}\\SVGD}
& \makecell{\textbf{Salt}\\Ours}
& \makecell{\textbf{Arid}\\SVGD}
& \makecell{\textbf{Arid}\\Ours}
\\ \midrule
L2 per sample $\downarrow$
& $4.5901 \pm 0.0207$ & \best{$0.6862 \pm 0.0915$}
& $7.3390 \pm 0.6018$ & \best{$6.3496 \pm 0.8867$}
& $2.9332 \pm 0.0087$ & \best{$2.1438 \pm 0.0583$}
\\
NRMS (\%) $\downarrow$
& $59.1007 \pm 0.3029$ & \best{$7.8003 \pm 0.9821$}
& $89.8641 \pm 8.8330$ & \best{$71.6360 \pm 10.7053$}
& $62.4468 \pm 0.3216$ & \best{$30.8195 \pm 1.5088$}
\\
Trace corr $\uparrow$
& $0.8410 \pm 0.0018$ & \best{$0.9965 \pm 0.0010$}
& $0.6564 \pm 0.0649$ & \best{$0.7560 \pm 0.0709$}
& $0.8476 \pm 0.0019$ & \best{$0.9352 \pm 0.0053$}
\\
\makecell{Mean $|\Delta t|$\\(ms) $\downarrow$}
& $3.2310 \pm 0.1038$ & \best{$0.0697 \pm 0.1196$}
& $17.6006 \pm 12.0422$ & \best{$4.3967 \pm 1.7985$}
& $4.3430 \pm 0.3252$ & \best{$2.5480 \pm 0.4443$}
\\
Envelope L1 $\downarrow$
& $3.4097 \pm 0.0229$ & \best{$0.3926 \pm 0.0127$}
& $5.6250 \pm 0.5704$ & \best{$4.4710 \pm 0.5432$}
& $1.8360 \pm 0.0053$ & \best{$0.8946 \pm 0.0232$}
\\
\makecell{Band spec\\relL2 $\downarrow$}
& $0.2192 \pm 0.0016$ & \best{$0.0124 \pm 0.0014$}
& $0.4308 \pm 0.0673$ & \best{$0.3615 \pm 0.0674$}
& $0.3591 \pm 0.0010$ & \best{$0.1387 \pm 0.0062$}
\\ \bottomrule
\end{tabular}
\end{table*}
\end{landscape}
\clearpage

\section{Discussions}

Utilizing the observed seismic shot gathers to do posterior sampling in FWI presents significant computational challenges. To mitigate this, we promote the use of a diffusion model-based posterior sampling algorithm utilizing the encoded-shot data. Although the encoded-shot data strategy has been widely recognized to reduce the required number of PDE solvers per FWI iteration, such a strategy is also associated with producing source-related artefacts courtesy of the blended data. To address this issue, we couple the use of such data with a diffusion model, which has been shown to regularize an FWI process. From a statistical point of view, we surrogate the prior distribution by training a diffusion model before doing posterior sampling. In this section, we continue the discussion of comparing our framework with SVGD before discussing its current limitations and potential solutions.


\subsection{Dependencies between posterior samples}

Apart from outperforming the SVGD algorithm as declared in the previous section, the proposed framework deviates from SVGD in its mechanism to produce posterior samples. Specifically, unlike SVGD, the proposed framework does not require interactions between posterior samples. In SVGD, each posterior sample interacts with each other through the Stein updates for each FWI iteration. Such a requirement translates into significant practical challenges when trying to do parallel posterior samples generation for large-scale 3D FWI experiments. In such scenarios, we have to take into account parallelization over source locations and posterior samples. In contrast, our framework treats each posterior sample independently, making it easier to do posterior sampling in parallel.

\subsection{The choice of stochastic refinement kernels}

The type of optimization algorithm used in this framework plays a significant role in ensuring high-quality posterior samples. This is because the choice of optimization algorithm in this work dictates the posterior sample exploration stage. In other words, a more accurate MCMC-type optimization algorithm will provide a better posterior sample quality. In contrast, such a preference will not be that influential in SVGD, as in this case, the evolution of posterior samples is governed by a deterministic transport (comprised of the log posterior (FWI) gradient and the radial basis function kernel). In this work, we consider studying the effect of a deterministic optimization algorithm as we focus on comparing our framework with SVGD. The potential issue that might arise from doing this is essentially the underestimation of variance due to a poor posterior sample exploration stage.

\subsection{Sampling efficiency}

Another potential improvement that we have not invested in is the choice of the diffusion sampler and architecture. We considered the use of a plain DDPM-style diffusion model for both of our 2D and 3D examples. While this is not the main focus of our work, we can further improve the sampling efficiency, particularly when handling large-scale 3D FWI by resorting to a more efficient diffusion model architecture (e.g., the latent diffusion model \cite{rombach2022high}). It is worth mentioning that while theoretically we can use a more efficient sampler than DDPM, e.g., the DDIM sampler \cite{song2020denoising}, careful selection of the noise weight is necessary to ensure a non-deterministic diffusion step.

\subsection{Other forms of guidance}

Finally, we base our experiment on the assumption that the observed seismic shot records are the only available information when doing posterior sampling. While this is quite a common scenario in the early-stage subsurface exploration, other forms of guidance might further improve the quality of the proposed framework. As shown in \cite{wang2024controllable, taufik2025efficient}, the diffusion model can admit other guidance conditions to further guide the posterior sampling process. Although incorporating a multi-modal guidance to our framework is trivial, a careful weighting strategy (between guidance modalities) is imperative to ensure meaningful posterior samples.

\section{Conclusions}


We introduced a diffusion–based posterior sampling framework for full waveform inversion (FWI) that leverages encoded (simultaneous–source) seismic data to reduce wave–equation solves while maintaining statistical fidelity. Encoded shots are known to introduce source–related crosstalk artifacts; in our approach, these artifacts are mitigated by the denoising capability of a diffusion prior, which simultaneously regularizes the model space during sampling. By combining (i) clean–space Langevin mixing at each diffusion level, (ii) patchwise re–noising (decoupled across levels), and (iii) unbiased mini–batch likelihoods from encoded shots, the proposed sampler (DAPS–style) consistently outperforms a conventional variational baseline (SVGD) across 2D synthetic and field experiments. In particular, we observe lower data misfit, reduced velocity–model error, and improved structural fidelity, at a substantially lower count of PDE solves per update.

Beyond accuracy and cost, the method is practical for large surveys: posterior draws are independent across chains, enabling straightforward parallelism over samples and over source locations. A key consideration, however, is the sensitivity of the sampler to the choice of the inner update: unlike SVGD—whose exploration is largely governed by the Stein kernel—the correctness and calibration of our method depend on using a stochastic Langevin step (with noise matched to any preconditioning). Deterministic optimizers (e.g., Adam without noise) turn the inner kernel into an optimizer and can lead to underestimated variance.

The computational cost remains dominated by wave–equation solves, especially in 3D; aggressive encoding or very small diffusion schedules can bias estimates if not calibrated. Performance depends on the realism of the learned prior and on accurate likelihood scaling for the supergathers. 

\section{Acknowledgement}

We thank King Abdullah University of Science and Technology (KAUST) and the DeepWave Consortium sponsors for their support. We also thank the Seismic Wave Analysis group for the supportive and encouraging environment. They are also grateful to the Seismic Laboratory for Imaging and Modeling Group for hosting the Compass model. This work utilized the resources of the Supercomputing Laboratory at KAUST in Thuwal, Saudi Arabia.

\vfill

\bibliographystyle{unsrt}  
\bibliography{references}

\end{document}